%% file: ms.tex
\input{definitions}

\documentclass{emulateapj}
\usepackage{natbib,psfig,amsmath}

\begin{document}
\shorttitle{Deep UV Luminosity Functions at the Infall Region of the Coma Cluster}
\shortauthors{Hammer et al.}
\title{Deep UV Luminosity Functions at the Infall Region of the Coma Cluster}
\author{D.~M.~Hammer$^{1,2}$,~A.~E.~Hornschemeier$^{2}$,~S.~Salim$^{3}$,~R.~Smith$^{4}$,~L.~Jenkins$^{2}$,~B.~Mobasher$^{5}$,~N.~Miller$^{6}$, and ~H.~Ferguson$^{7}$}
\altaffiltext{1}{Department of Physics and Astronomy, The Johns Hopkins University, 3400 N. Charles Street, Baltimore, MD 21218, USA}
\altaffiltext{2}{Laboratory for X-ray Astrophysics, NASA GSFC, Code 662.0, Greenbelt, MD 20771, USA}
\altaffiltext{3}{Department of Astronomy, Indiana University, Bloomington, IN 47404, USA}
\altaffiltext{4}{Department of Physics, University of Durham, South Road, Durham DH1 3LE, UK.}
\altaffiltext{5}{Department of Physics and Astronomy, University of California, Riverside, CA 92521, USA}
\altaffiltext{6}{Department of Astronomy, University of Maryland, College Park, MD, 20742-2421, USA}
\altaffiltext{7}{Space Telescope Science Institute, 3700 San Martin Drive, Baltimore, MD 21218, USA}

\begin{abstract}
We have used deep {\it GALEX} observations at the infall region of the Coma cluster to measure the faintest UV luminosity
functions (LFs) presented for a rich galaxy cluster thus far. The Coma UV LFs are measured to M$_{\mathrm{UV}}$ $=-10.5$ in
the {\it GALEX} FUV and NUV bands, or 3.5 mag fainter than previous studies, and reach the
dwarf early-type galaxy population in Coma for the first time.
The Schechter faint-end slopes
($\alpha\approx-1.39$ in both {\it GALEX} bands) are shallower than reported in previous Coma UV LF studies
owing to a flatter LF at faint magnitudes.
A Gaussian-plus-Schechter model provides a slightly better parametrization of the UV LFs resulting in a faint-end slope of $\alpha$ $\approx$ $-1.15$ in both {\it GALEX} bands.
The two-component model gives faint-end slopes shallower than $\alpha=-1$ (a turnover) for the LFs constructed separately for passive and star forming galaxies.
The UV LFs for star forming galaxies show a turnover at M$_{\mathrm{UV}}$ $\approx$ $-14$ owing
to a deficit of dwarf star forming galaxies in Coma with stellar masses below M$_{*}=10^{8}~\msun$.
A similar turnover is identified in recent UV LFs measured for the Virgo cluster suggesting this may be a common feature of local galaxy clusters,
whereas the field UV LFs continue to rise at faint magnitudes.
We did not identify an excess of passive galaxies as would be expected if the missing dwarf star forming galaxies were
quenched inside the cluster.
In fact, the LFs for both dwarf passive and star forming galaxies show the same turnover at faint magnitudes.
We discuss the possible origin of the missing dwarf star forming galaxies in Coma and their expected properties based on 
comparisons to local field galaxies.
\end{abstract}

\pagebreak

\section{Introduction}
The {\it nature} vs.~{\it nurture} debate is a fundamental theme of galaxy evolution studies regarding the relative importance of internal and external processes in the
star formation histories of galaxies. Ultraviolet (UV) observations of galaxies are critical to this debate as they provide a direct measure of star formation by 
probing short-lived ($\simlt$1 Gyr) massive stars \citep[e.g.,][]{Kennicutt1998}.  For example, the rest-frame UV luminosity function (LF) has allowed measurements of the cosmic
star formation rate at high redshifts \citep[1$\simlt$$z$$\simlt$9; e.g.,][]{Madau1996,Bouwens2010},
and space-based UV facilities such as {\it GALEX} have confirmed the rapid decline of star formation from $z\sim1$ to the present day \citep[e.g.,][]{Arnouts2005,Schiminovich2005}.
Environment may play an important role in this decline as the galaxies that dominate the star formation density at $z\sim1$ tend to be located in dense regions \citep[e.g.,][]{Cooper2006, Bell2007}
and their star formation rates have declined at a relatively faster rate to the present day \citep{Li2011}.
Moreover, models suggest that a large fraction of these galaxies now reside in local groups and clusters \citep{Crain2009,Cen2011}.
UV LF studies of local galaxy clusters may 
therefore provide important constraints on the processes that regulate star formation in dense environments, 
giving us a local proxy for studying these effects in the distant Universe.
There is considerable debate, however, regarding the identity of the dominant processes that operate in cluster environments \citep[see the review by][]{Boselli2006a}.
For normal late-type galaxies, the removal of gas and the subsequent quenching of star formation within infalling cluster members is typically
attributed to interactions with the hot intracluster medium
\citep[ICM; e.g.,``ram-pressure stripping", ``starvation" or ``strangulation", ``thermal evaporation";][]{Gunn1972,Cowie1977,Larson1980,Bekki2002}.
Tidal processes are invoked to explain the transformation to early-type morphology such as interactions with the cluster potential \citep[e.g.,][]{Merritt1983,Byrd1990},
and/or encounters with other galaxies \citep[e.g., ``harassment", ``tidal stirring";][]{Moore1996,Moore1999,Mayer2001}.

Dwarf galaxies are potentially key to investigating this issue, as
their shallow potential wells make them more susceptible to cluster
processes. Moreover, models suggest that while giant galaxies are
typically accreted into clusters as members of massive groups, and
hence to have been ``pre-processed" in the group environment, dwarf galaxies
are more likely to have been accreted via low-mass halos \citep[e.g.,][]{McGee2009}.
Hence the impact of the cluster should be more cleanly distinguished for dwarf galaxies.


We lack a clear understanding of the evolution of dwarf galaxies and the origin
of the abundance of dwarf early-type (dE) galaxies found in local clusters \citep[e.g.,][]{Ferguson1994,Jenkins2007,Weinmann2011}.
This is an important endeavor as dE galaxies dominate the faint-end slopes of cluster LFs
which constrain $\Lambda$CDM models of galaxy formation and evolution \citep[e.g.,][]{Benson2003LFb,Guo2011}.
dE galaxies may result from recent quenching of infalling dwarf galaxies owing to processes that are active from the periphery of the cluster
to the core, and possibly with galaxy groups prior to accretion
\citep[e.g.,][]{Treu2003,Pogg2004ICM,Moran2007,LiskerIII}.
On the other hand, some dE galaxies may not be linked to recent quenching of the infall population,
e.g., dE galaxies may survive multiple orbits inside the cluster \citep[e.g.,][]{Tully2002,Pimbblet2011} or may form
from the tidal debris of massive cluster members \citep{Duc2007}.

Coma is one of the nearest examples of a rich galaxy cluster and its large dwarf galaxy
population is ideal for studying the effects of external processes on star formation.
Previous studies in Coma have identified the signatures of dynamical processes 
associated with both the ICM \citep[e.g.,][]{Caldwell1993,Bravo2000,Castander2001,Fino2004LF, Pogg2004ICM,Gavazzi2006,Miller2009a,Smith2010,Yagi2010}
and tidal interactions \citep[][]{Thompson1981,Graham2003,Marinova2010,Madrid2010,Chiboucas2010,Peng2011}.
Most of these studies were based on small galaxy samples or case studies of unique disrupted populations such as post-starburst galaxies, stripped disks, 
barred galaxies, ultracompact dwarf galaxies (UCDs), and intracluster globular clusters.

In this paper, we present deep UV LFs for the Coma cluster to study the environmental impact on the star formation properties of the aggregate dwarf galaxy population.
Our field is located between the Coma center and the projected virial radius so that we may observe
galaxies affected by the environment but before severe tidal disruption at the core.
Moreover, the accretion histories of dwarf star forming and dE galaxies are better matched at the infall region as compared to the center 
\citep[e.g.,][]{Cortese2008b,Smith2011}, which allows a more straightforward comparison of their LFs in order to constrain models of dE galaxy formation.
UV LFs present advantages over optical studies for this purpose as UV data allows a more reliable separation of star forming and passive galaxies,
especially for dwarf galaxies \citep{Haines2008}.


Previous Coma LFs \citep[a complete compilation is provided in][]{Jenkins2007} have been measured in bands that trace star formation such as UV \citep{Donas1991,Andreon1999,Cortese2003a,Cortese2003b,Cortese2008b},
H$\alpha$ \citep{Iglesias2002}, optically-selected blue galaxies \citep{Mobasher2003LF}, 24$\mu$m \citep{Bai2006}, and radio wavelengths \citep{Miller2009b}.
Interestingly, most studies have found little or no difference of the faint-end slope for star forming galaxies with respect to the field environment, other local clusters, or 
with clustercentric distance. We expect a turnover for dwarf star forming galaxies as they should be more susceptible to dynamical processes.

Here, we present the UV LFs of dwarf star forming galaxies to lower masses than these other studies in an attempt to identify such a turnover, and we include the 
passive dwarf galaxy population in Coma for the first time in order to constrain models of dE galaxy formation.
We make use of deep 26 ks {\it GALEX} FUV and NUV observations of the Coma infall region with SDSS photometric coverage and moderately deep redshift coverage ($r$$\sim$21.2) to extend
the faint end of the Coma UV LF to 3.5 magnitudes deeper than previous efforts. We cover depths similar to the faintest {\it GALEX} LFs presented thus far for the (much closer) Virgo cluster 
\citep{Boselli2011}.
We assume Coma is located at a distance of 100 Mpc \citep[e.g.,][]{Carter2008} which corresponds to a distance modulus of DM $=35.00$ mag ($z=0.023$)
and angular scale of 1.667 Mpc deg$^{-1}$ for H$_{0}=71$~km s$^{-1}$ Mpc$^{-1}$, $\Omega_{\lambda}=0.73$, and $\Omega_{m}=0.27$.

\section{Data}
\subsection{GALEX and SDSS Catalog}
In \citet[][hereafter Paper I]{Hammer2010a}, we describe our {\it GALEX} observations in detail, and the construction of a source catalog for $\sim$9700 galaxies with {\it GALEX} and SDSS photometry.
Here we provide only a brief summary.
The field is located 0\fdg9 (1.5 Mpc) south-west of the Coma center and inside the cluster virial radius \citep[r$_{vir}$=2.9 Mpc;][]{Lokas2003}.
The spatial location of our {\it GALEX} field is shown in Figure \ref{coma_field}.
Photometry was taken from the {\it GALEX} pipeline catalog for bright (m$_{\mathrm{UV}}<21$) or extended galaxies,
where extended galaxies are defined as having a 90\% flux radius larger than 10$\arcsec$ in the SDSS $r$-band.
{\it GALEX} pipeline apertures were inspected for accuracy and custom photometry was performed in cases of shredded or blended objects.
A PSF deblending technique \citep[][]{Guillaume2006} was used to measure faint and compact sources which lessens systematic effects from object blends, source confusion, and the Eddington Bias, resulting
in significantly improved detection efficiency and photometric accuracy as compared to the pipeline catalog.
This hybrid approach was necessary given the nature of our very deep images that consist of many resolved cluster members superposed on a dense field of unresolved background galaxies.

The catalog has a limiting depth of NUV=24.5 and FUV=25.0 mag in the central 0\fdg5 {\it GALEX} field-of-view (FOV).
The {\it GALEX} images extend over a larger area but we restrict faint sources to the inner region where the PSF is well behaved
as required by the deblending method.
We choose a magnitude limit of m$_{\mathrm{UV}}=24.5$ for both {\it GALEX} bands which corresponds to the 95\% completeness limit of the {\it GALEX} images.
The coverage area is larger for galaxies brighter than m$_{\mathrm{UV}}=21$ mag (the inner 0\fdg6 {\it GALEX} FOV)
as the pipeline aperture photometry is not affected by the shape of the PSF.
The larger coverage area increases the number of bright cluster members in our sample which allows for better constraints on the bright-end turnover of the UV LFs.
UV magnitudes are corrected for foreground Galactic extinction and are reported in the AB system.
Typical photometric uncertainties span 0.1$-$0.3 mag (1$\sigma$ rms scatter) for bright and faint galaxies, respectively, in both {\it GALEX} bands.

We rely on the SDSS data to identify galaxies in our catalog as {\it GALEX} lacks the spatial resolution to perform this task.
The SDSS counterparts to {\it GALEX} sources extend to optical magnitudes of $r$=24.
For this study, however, we have imposed an optical cutoff at $r$=21.2 mag for the following reasons:
(1) spectroscopic redshift coverage in our {\it GALEX} field is incomplete at fainter magnitudes,
(2) SDSS photometric completeness drops rapidly at fainter magnitudes (see Appendix), and
(3) SDSS star/galaxy classification is less reliable at fainter magnitudes (see Appendix in Paper I).
As a result of this optical cutoff, our sample does not include potential Coma member galaxies that are fainter than $r$=21.2
mag, but are brighter than our UV magnitude limits at m$_{\mathrm{UV}}$=24.5 (i.e.,~optically-faint galaxies with very blue UV-optical
colors). Galaxies with colors redder than m$_{\mathrm{UV}}$$-r$$=$3.3, which includes the entire cluster red sequence,
are not affected and have full coverage to our chosen UV magnitude limits.

The photometric completeness of our catalog is addressed in the Appendix with particular attention given to the SDSS
detection efficiency; the {\it GALEX} detection efficiency is better than 95\% across the full UV magnitude range (Paper I).
The SDSS detection efficiency was measured directly by comparing DR6 detections to optical source catalogs
from the much deeper and higher-resolution HST-ACS images at the center of Coma \citep{Hammer2010b}.
This comparison was performed separately for both background galaxies and Coma members.
The SDSS detection efficiency for Coma members is near 100\% to $r=20.5$, and drops to $\sim$70\% at our adopted magnitude limit
of $r=21.2$. The detection rates are slightly higher for background galaxies at a given apparent magnitude 
as they tend to have higher surface brightness than the dwarf galaxy population in Coma.
The SDSS detection efficiency has also been converted to UV magnitudes for the 
galaxies in our sample. The SDSS detection rates are given in tabular form in the Appendix.

Our sample consists of 3125 galaxies after removing objects fainter than $r$=21.2 mag.
We describe the spectroscopic redshift coverage in our field in Section 2.2 and
remove obvious background galaxies from our galaxy sample in Section 2.3. A summary of the final catalog is given in Section 2.4.


\subsection{Spectroscopic Redshift Coverage}
\begin{figure}[t!]
\vspace{50pt}
\plotone{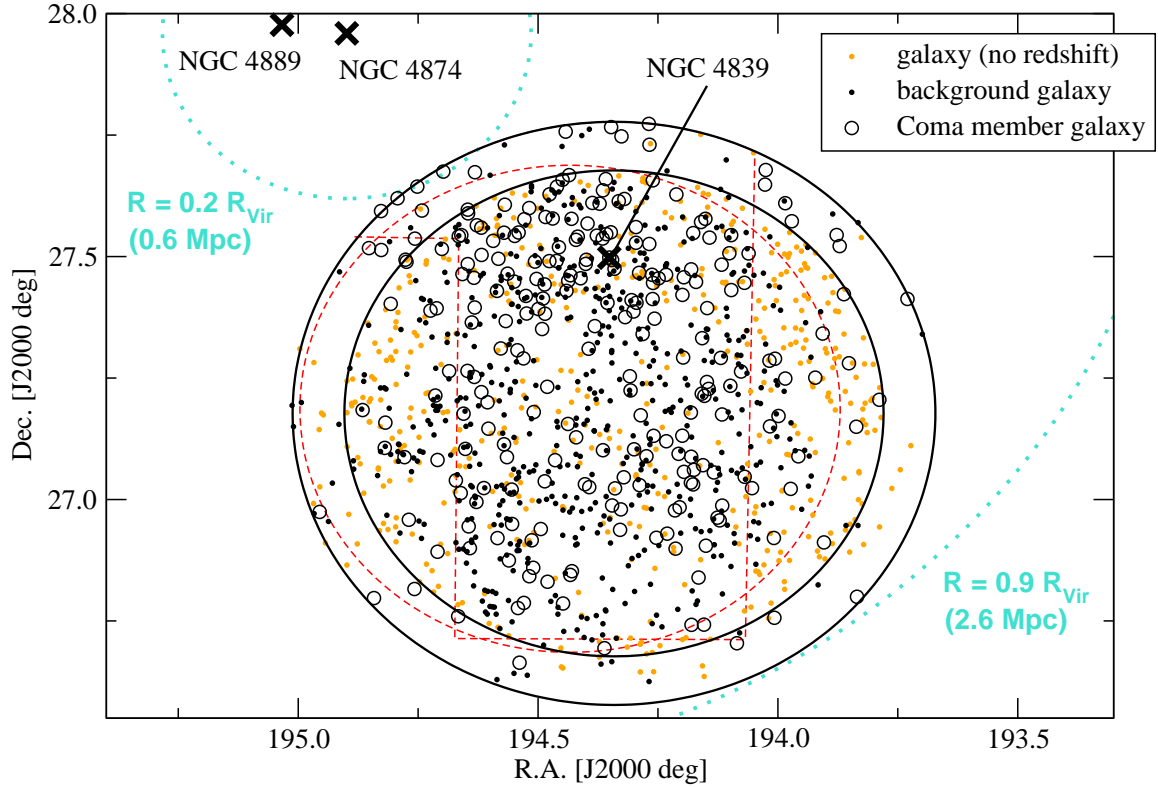}
\caption{\label{coma_field} Location of our {\it GALEX} field at the Coma-3 infall region of the Coma cluster.
Bold Xs mark the three largest galaxies in Coma (NGC 4889, 4874, 4839).
Concentric circles (black) show the inner 0\fdg5 and 0\fdg6 {\it GALEX} FOV which span
cluster-centric distances of R = 0.2 and 0.9 R$_{\mathrm{vir}}$ (blue dotted lines).
{\it GALEX} coverage extends to NUV=24.5 and FUV=24.5 mag inside the
inner circle, and NUV=21.0 and FUV=21.0 mag inside the outer circle.
Spatial locations are shown for the 1297 galaxies in our trimmed catalog (see Section 2.4), separated 
into three subsets that include galaxies that lack redshifts (458 orange dots), confirmed 
background galaxies (591 black dots), and confirmed Coma member galaxies (248 open black circles).
SDSS photometric and spectroscopic coverage is available across the full {\it GALEX} FOV.
Deeper spectroscopic redshift coverage is available from surveys performed with MMT-Hectospec (dashed red circle) 
and WHT \citep[dashed red rectangular regions;][]{Mobasher2001spectra}.}
\end{figure}

Several optical spectroscopic redshift campaigns have been conducted in the region of Coma studied here.
These campaigns include surveys that span the whole {\it GALEX} FOV such as SDSS DR7 \citep[$r$$<$17.77;][]{Adelman2008}
and the spectroscopic follow-up survey of the GMP \citep*{Godwin1983} optical photometric catalog \citep*[B$_{j}$$<$20;][]{Colless1996}.
Smaller studies provide additional redshift measurements as compiled in NED \cite[e.g.,][]{Kent1982, Caldwell1993,vanHaarlem1993,Biviano1995,Colless01,Rines2003}.
A deep spectroscopic survey was performed with the William Hershel Telescope that covers $\sim$66\% of the inner 0\fdg5 {\it GALEX} FOV and a smaller
fraction of the outer field as seen in Figure \ref{coma_field} \citep[1$<$$B$-$R$$<$2 and $R$$<$20;][]{Mobasher2001spectra}. There is a more recent
spectroscopic campaign with MMT-Hectospec that provides deeper redshift coverage ($\sim1$ hour exposures) to
$r$$\sim$21.5 mag within our field and the cluster center. The MMT redshift targets ($\sim$6300 total) were selected from the
SDSS DR6 photometric catalog with an emphasis on objects fainter than the SDSS spectroscopic limit at $r=17.77$.
The targets were selected with minimal surface brightness and optical color bias, especially for the region of the cluster studied here (a subset of redshift targets
at the center of the cluster were limited to colors bluer than $g-r=1.2$). As shown in Figure \ref{coma_field}, the MMT survey covers the majority of the inner 0\fdg5
{\it GALEX} FOV with partial coverage of the western outer field. A brief description of the MMT main redshift survey in Coma is given in several studies
\citep{Smith2008,Smith2009a,Chiboucas2010}.

A composite redshift catalog was created from these spectroscopic surveys that consists of 2483 redshifts in the field studied here.
The majority of redshifts are taken from the SDSS and MMT surveys for galaxies brighter/fainter than $r=17.77$, respectively.
We matched 1707/2483 redshifts to the 3125 galaxies in our photometric catalog described in Section 2.1.
The unmatched redshifts (776/2483) are either stars or galaxies that are fainter than our chosen UV or $r$-band magnitude limits
x(most unmatched redshifts are located in the outer $0\fdg5-0\fdg6$ {\it GALEX} FOV where we impose bright magnitude limits).

We also use the partial HST-ACS coverage of our field \citep[$\simlt$2\%;][]{Carter2008} to determine membership for 8 additional galaxies based on morphology \citep[see][]{Chiboucas2010}.
We identified 5 background galaxies and 3 Coma members among the galaxies in our sample that lack redshift coverage. We consider these 8 galaxies as part of our spectroscopic sample
for the remainder of the paper.

\subsection{Excluding Obvious Background Galaxies}
We must be aware of potential selection effects in the redshift surveys that would bias measurements of the LF.
Some redshift surveys, for example, have specifically targeted cluster member galaxies via color selection.
In the opposite sense, spectroscopic surveys that select targets based only on apparent magnitude may have poorer coverage of faint cluster member galaxies,
i.e., dwarf galaxies tend to have lower surface brightness (thus lower S/N spectra) as compared to background
galaxies for a given apparent magnitude.

\begin{figure}[t!]
\vspace{50pt}
\plotone{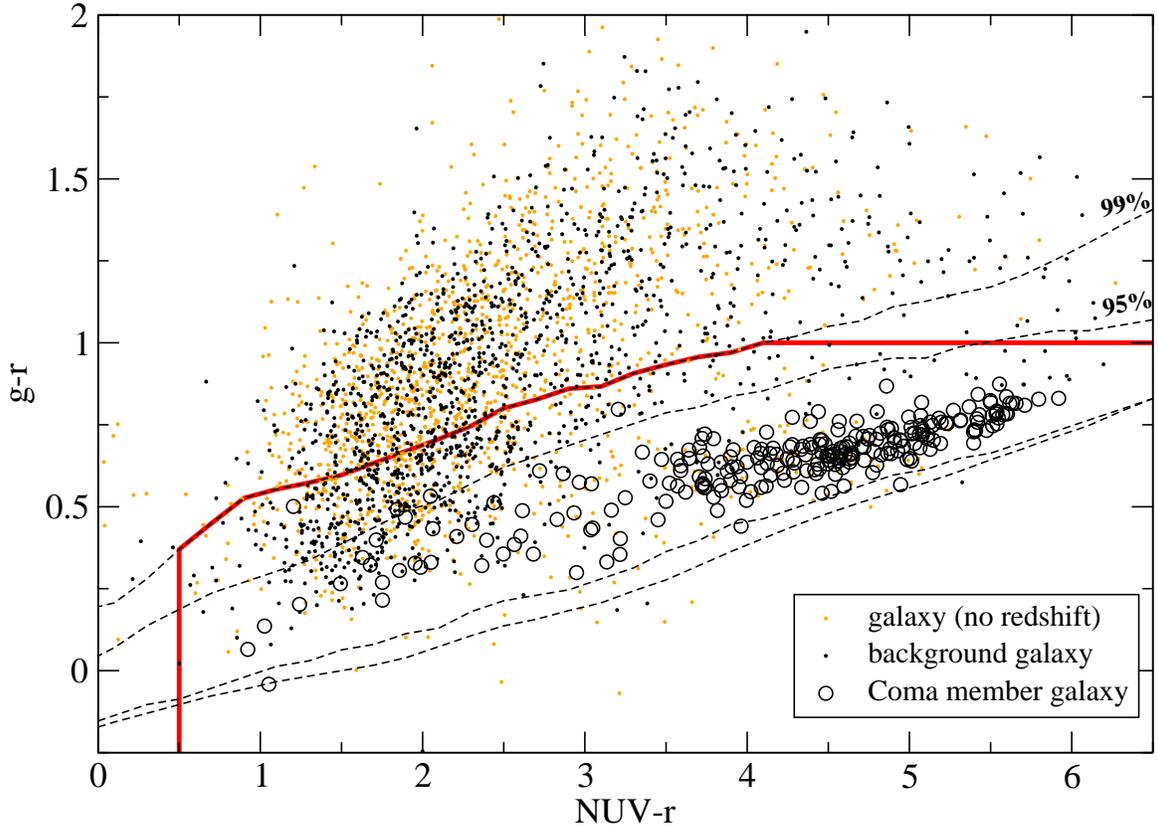}
\caption{\label{uvopt} UV-optical color-color diagram for galaxies in the full {\it GALEX}/SDSS catalog.
The sample is separated into three subsets that include galaxies that lack spectroscopic redshifts (orange dots), spectroscopically confirmed
background galaxies (black dots), and confirmed Coma member galaxies (open black circles).
Dashed lines trace the bi-variate density distribution of $\sim$100,000 model SEDs \citep[from][see Section 2.3 for details]{Salim2007} that 
are used to identify the color space where Coma member galaxies are expected; the inner
and outer contours enclose 95\% and 99\% of the models, respectively.
We use the thick red line to exclude 1828 obvious background galaxies that are shifted above this cutoff owing to K-corrections and keep the
1297 galaxies located below the line. The red sequence of cluster member galaxies is clearly visible at UV-optical colors redder than NUV-$r\approx3.5$.}
\end{figure}

We can mitigate these potential selection effects by excluding galaxies at much higher redshifts that tend to have both redder optical colors and higher surface brightness
than galaxies in the Coma cluster.
In Figure \ref{uvopt}, a UV-optical color-color diagram shows the location of all galaxies in our catalog including both confirmed Coma
member galaxies and background galaxies. We identify the region of the diagram where we expect Coma member galaxies using
a Monte Carlo suite of multiple-burst SEDs ($\approx$100,000 models) that span a large range in metallicity, star formation history, and
dust attenuation as described in \cite{Salim2007};  the contours in Figure \ref{uvopt} show the regions of the diagram
covered by 95\% and 99\% of the models.
The galaxies located above the models are objects at higher redshifts ($z$$\simgt$0.2) that are shifted into this region owing to large $K$-corrections.
We have removed the 1828 galaxies that lie above the thick line shown in Figure \ref{uvopt}, which was chosen conservatively to include all potential Coma members and to roughly follow the models.
We retain the 1297 galaxies located below the cutoff.

\subsection{Catalog Summary}
The final catalog used in this study consists of 1297 objects that have {\it GALEX} and SDSS detections brighter than NUV=24.5 and $r$=21.2 mag, respectively.
A subset (1193/1297 or 92\%) have FUV detections brighter than our chosen limit at FUV$=$24.5 mag.
Galaxies were identified via the SDSS star/galaxy classification.
We have restricted our catalog to the region of $g$-$r$ vs.~NUV-$r$ color space where we expect galaxies at the distance of Coma.
Spectroscopic redshifts are available for 839 galaxies (65\%) including 248 members of the Coma cluster \citep[4000$<$v$<$10 000 km s$^{-1}$;][]{Colless1996}.
The spatial locations of our galaxies are shown in Figure \ref{coma_field}.
An analysis of the photometric completeness for our catalog is provided in the Appendix.

\section{UV and Optical Properties of Coma Members}
In this section, we use UV-optical color diagnostics to classify the Coma galaxy population according to star formation activity.
This separation is necessary for reliable measurements of the UV LF (Section 4),
and allows us to assess the LFs for different galaxy populations (star forming and passive).
We classify dwarf galaxies in Coma by optical magnitude ($r\geq16.5$ or M$_{r}\geq-18.5$) which corresponds to
a Johnson B--band magnitude of M$_{\mathrm{B}}\sim-17.0$ \citep[most studies adopt a cutoff between --18 $<$ M$_{\mathrm{B}}<-16$;][]{Ferguson1994}.


\subsection{FUV$-r$ Color vs.~Specific Star Formation Rate}
The FUV band allows for a clean separation of star forming and passive galaxies as it is sensitive to recent star formation
(a timescale of roughly 0.1 Gyr as compared to 1 Gyr for the NUV band).
As such, we use FUV$-r$ color as a proxy for the specific star formation rate (SSFR) with the $r$-band roughly tracing the stellar mass.
We note that NUV$-r$ color is a slightly more reliable indicator of SSFR for star forming galaxies as it is less affected by dust,
but its sensitivity to metallicity does not allow for a clean separation of metal-poor passive galaxies and star forming galaxies using a simple color cut.

\begin{figure}[t!]
\vspace{50pt}
\centerline{\includegraphics[angle=0.,scale=1]{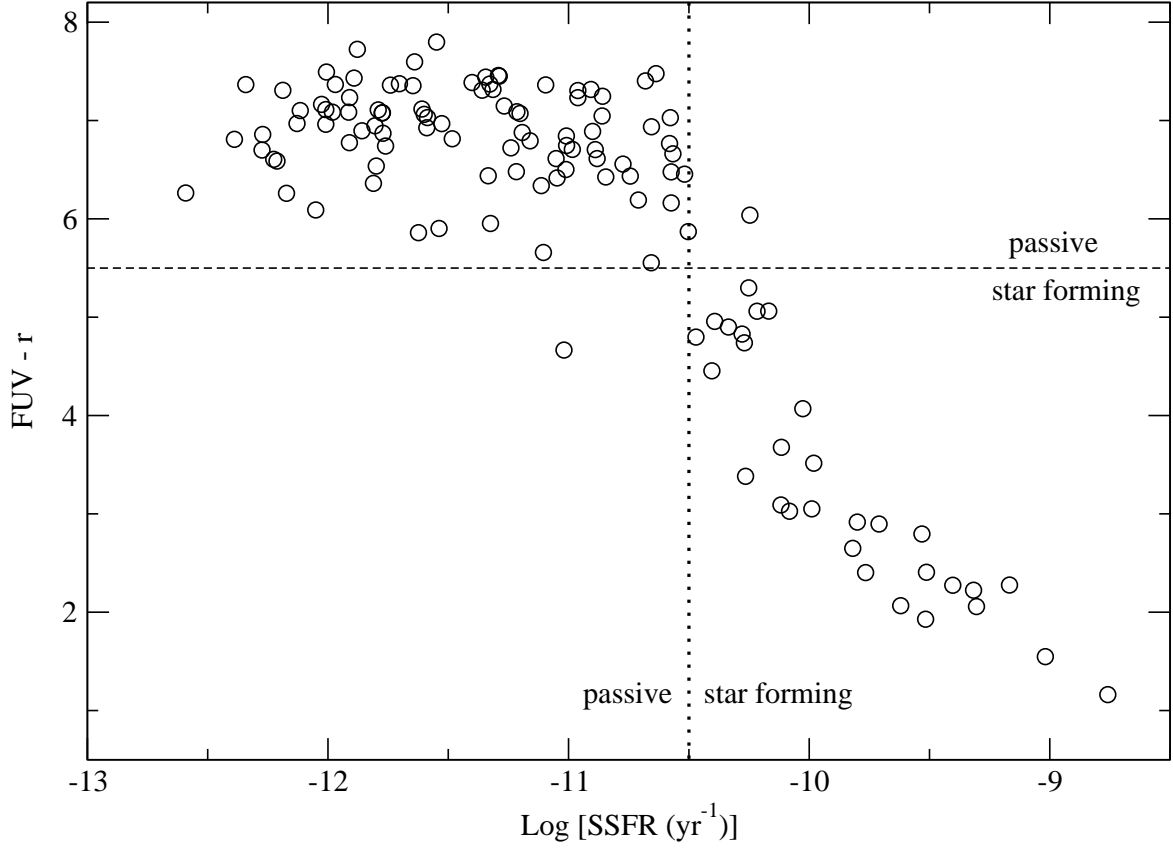}}
\caption{\label{ssfr} FUV$-r$ color vs.~specific star formation rate (SSFR) for Coma member galaxies.
SSFRs are taken from the MPA-JHU SDSS value-added catalog \citep[e.g.,][]{Brinchmann2004} which is limited to galaxies brighter than $r$$\approx$17.8 mag.
The vertical dotted line indicates the typical separation in SSFR for galaxies classified as star forming or passive \citep[e.g.,][]{Salim2007}.
The diagram shows that FUV$-r$ color allows for a similar separation of SF and passive galaxies in Coma with a dividing line at FUV-r=5.5 mag (horizontal dashed line).}
\end{figure}

In Figure \ref{ssfr}, we compare the FUV$-r$ colors of Coma member galaxies to SSFRs taken from the MPA-JHU value-added catalog which are measured via SDSS DR7 spectroscopic emission lines 
\citep[][]{Brinchmann2004}. The MPA-JHU values are corrected for light outside the SDSS 3$\arcsec$ fiber by fitting the SDSS photometry to stochastic SED models.
For non-SF (passive) galaxies, the MPA-JHU SSFRs are estimated from the 4000 \AA~break \citep[][]{Brinchmann2004}, and the aperture corrections from SED fitting 
now account for previous overestimates of their SSFRs as identified by \cite{Salim2007}.
The FUV$-r$ colors are not corrected for internal dust attenuation whereas the SSFRs are corrected for internal dust and, in some cases, for AGN emission.
The diagram identifies the SSFRs that correspond to star forming and passive galaxies \citep[taken from][]{Salim2007}.
We are able to select star forming galaxies (Log SSFR (yr$^{-1}$) $>$ $-10.5$) using a simple color cut below FUV$-r$ $=5.5$ with few outliers.
A separation based on the observed colors is necessary as we do not have reliable dust indicators for all galaxies in our sample.

For passive galaxies, the FUV$-r$ colors result from a mixture of FUV sources in each galaxy, such as possible AGN emission and/or 
stars from residual star formation, post-AGB stars, and low-mass helium-burning stars on the horizontal branch (HB; e.g.~extreme HB stars or blue HB stars).

\subsection{UV$-r$ vs.~$r$ CMDs}
\begin{figure}[t!]
\vspace{50pt}
\centerline{\includegraphics[angle=0.,scale=1]{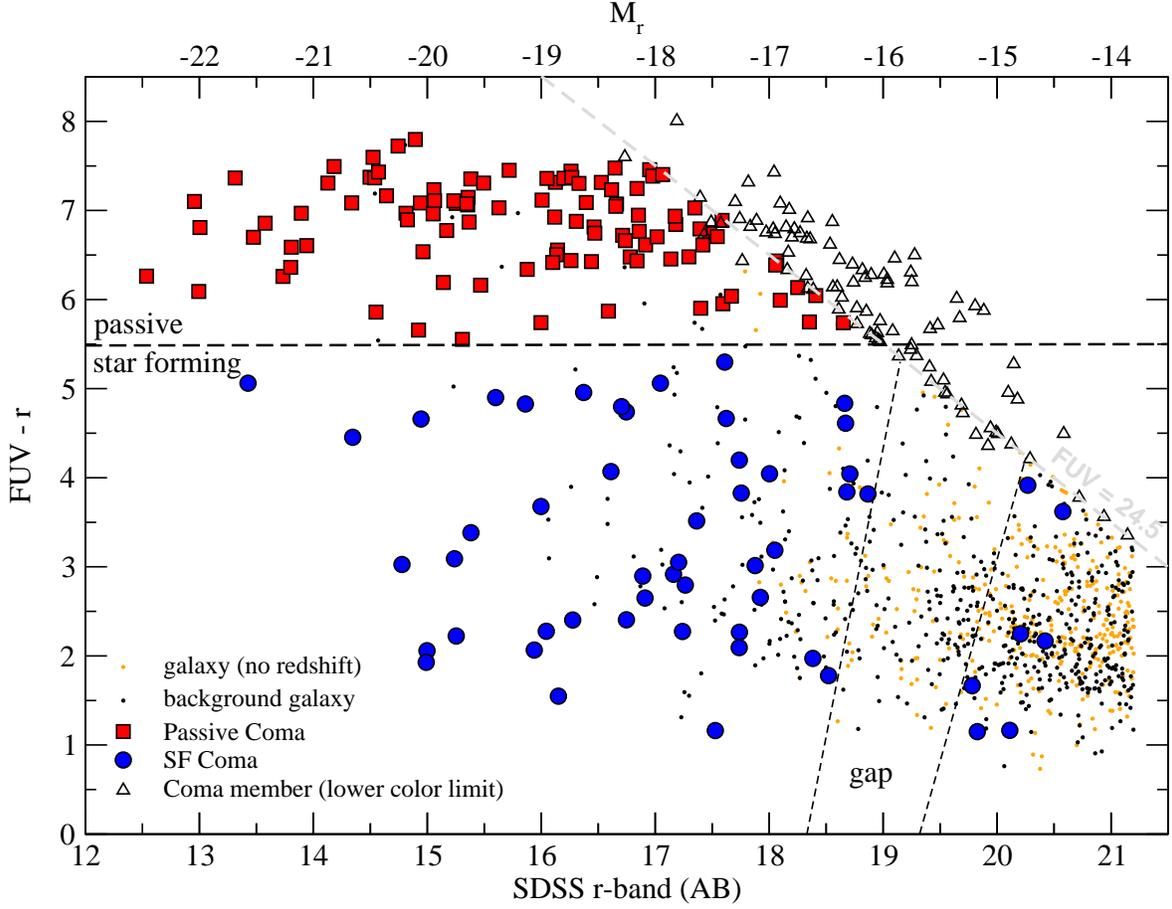}}
\caption{\label{fuvr} FUV-optical CMD for all galaxies in our sample. Orange dots are galaxies that lack spectroscopic redshifts, 
black dots are confirmed background galaxies, and confirmed Coma members are shown as large filled symbols (red squares=passive; blue circles=star forming).
We have spectroscopic redshift coverage across the full magnitude-color space.
We use the horizontal dashed line at FUV$-r=5.5$ to separate star forming and passive galaxies in the Coma cluster.
The top x-axis gives the corresponding absolute $r$-band magnitude for galaxies at the distance of Coma.
Triangles show the lower color limits for Coma member galaxies that are detected in the NUV band but have FUV magnitudes fainter than our chosen limit at FUV$=24.5$ (gray dashed line);
the lower color limits are used to classify those Coma members above FUV$-r=5.5$ as passive.
The black dashed lines define a ``gap" in the CMD that is void of confirmed Coma member galaxies (see Section 3.3).}
\end{figure}

The FUV$-r$ vs.~$r$ CMD for our catalog is presented in Figure \ref{fuvr}.
The diagram separates the galaxies by redshift classification: those galaxies that lack redshifts, spectroscopically confirmed background galaxies, and
spectroscopically confirmed Coma members (star forming and passive).
Background galaxies are more prevalent at blue colors and are the majority population at $r$$\simgt$18,
whereas cluster members are the majority at brighter optical magnitudes where the redshift coverage is nearly complete.
The redshift coverage spans the full color space.
The cluster red sequence is the horizontal band of passive galaxies at the top of the diagram.
Several Coma members are fainter than the FUV magnitude limit at FUV=24.5 but are brighter than NUV=24.5 and therefore included in our catalog.
We have measured their lower FUV$-r$ color limits\footnote{FUV lower limits were measured on the background-subtracted
{\it GALEX} pipeline images inside a circular aperture covering the FWHM PSF. The flux was scaled by a factor of 2 (the galaxies are nearly point sources)
and corrected for Galactic extinction. Magnitudes were taken at the lower 2$\sigma$ flux value (95\% confidence interval) or set to FUV=24.5 if photometry failed.}
which allowed us to classify additional passive galaxies.

\begin{figure}[t!]
\vspace{40pt}
\centerline{\includegraphics[angle=0.,scale=1]{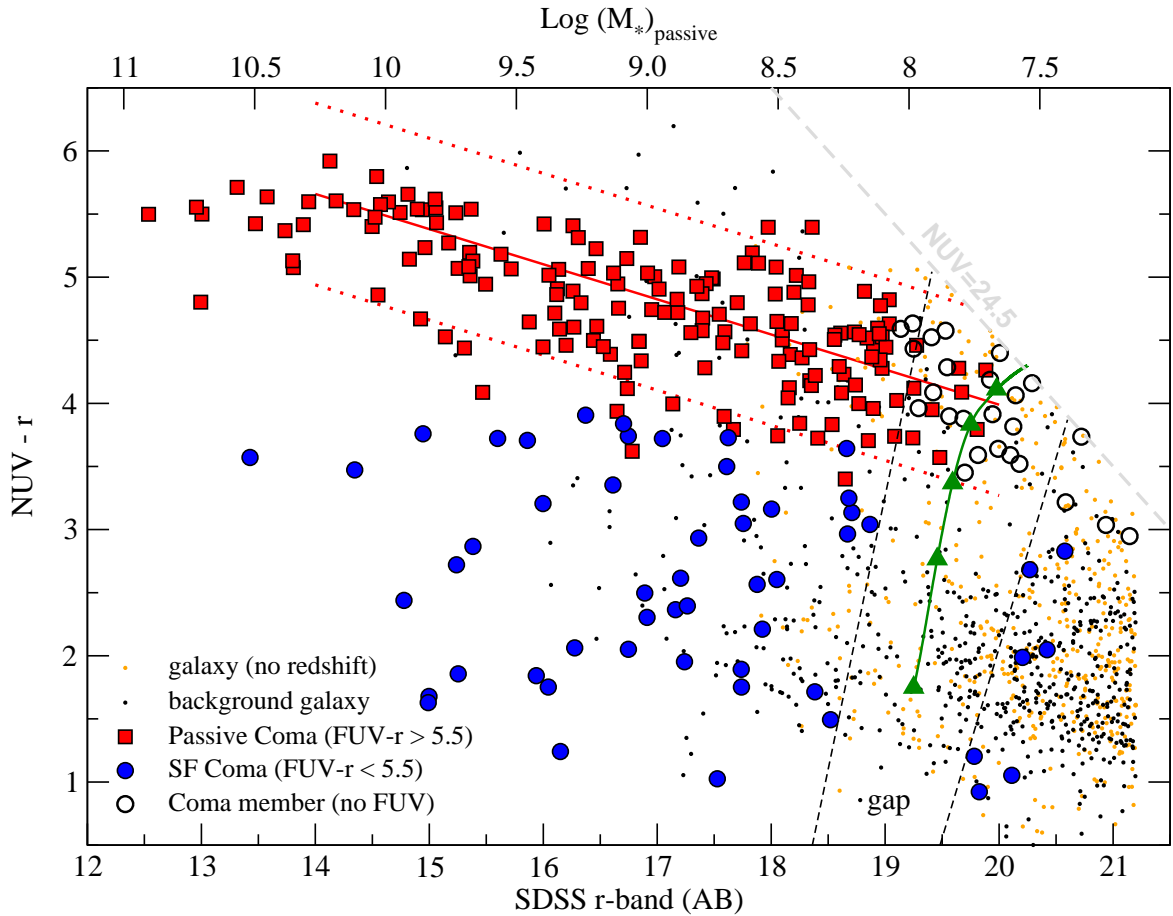}}
\caption{\label{nuvr} NUV-optical CMD for all galaxies in our sample. Orange dots are galaxies that lack spectroscopic redshifts, 
black dots are confirmed background galaxies, and confirmed Coma members are shown as large symbols (red squares=passive;
blue circles=star forming; black open circles=no FUV color information).
We have spectroscopic redshift coverage across the full magnitude-color space.
A linear fit to passive Coma member galaxies (red solid line) and the 2$\sigma$ scatter (red dotted lines) are shown.
The majority of Coma members without FUV colors are passive galaxies as they are located inside the 2$\sigma$ scatter.
The top x-axis gives the average stellar masses for passive Coma member galaxies as measured in this study (Log M$_{*}$ $= 16.364 - 0.435$$\times$$r$; Section 3.2);
SF galaxies are $\sim$0.5 mag brighter in the $r$-band for a given stellar mass.
The black dashed lines enclose a ``gap" in the CMD that is void of confirmed Coma member galaxies. Inside the gap, the thick green 
line is the path that a dwarf SF galaxy in Coma would follow after a rapid 100 Myr quenching event (see Section 3.3);
the line spans 4 Gyr from the onset of quenching and triangles show the model after 0, 0.25, 0.5, 1, and 2 Gyr.
The gray long-dashed line is our chosen magnitude limit at NUV=24.5.}
\end{figure}

The NUV$-r$ vs.~$r$ CMD is shown in Figure \ref{nuvr} with the same symbols as used for the FUV CMD.
The redshift coverage spans the entire NUV CMD color space.
Passive Coma members trace bluer colors at faint optical magnitudes as opposed the flat red sequence seen in the FUV CMD.
We performed a linear $\chi$$^{2}$ fit to the red sequence (0.36 mag scatter) as shown in Figure \ref{nuvr}.
The Coma member galaxies that lack FUV colors for SF/passive classification are indicated on the diagram.
The majority are passive galaxies as they are located inside the 2$\sigma$ fitted red sequence (the 3 unclassified cluster members at NUV$-r$$<$3.2 are considered SF galaxies).

The tilt of the red sequence in the NUV CMD is likely a metallicity-mass effect owing to NUV emission that is dominated by main-sequence turnoff stars \citep[MSTO; e.g.,][]{Dorman2003},
i.e., for a given stellar age, metal-poor (low-mass and optically-faint) passive galaxies have a more massive MSTO with brighter NUV emission.
The average stellar ages of passive galaxies tend to be younger for lower mass galaxies so part of the tilt also reflects an age-mass relation \citep*[e.g.,][and many others]{Caldwell2003,Michielsen2008,Smith2009b}.
Emission from MSTO stars does not extend appreciably into the FUV band resulting in its relatively flat red sequence.
Optically-bright galaxies on the red sequence ($r$$<$14) have slightly bluer colors as compared to the fitted relation.
These may be UV-upturn galaxies (also visible in the FUV CMD) that are brighter in UV owing to EHB stars \citep[e.g.,][]{Salim2010,Carter2011}.


The average stellar masses of passive galaxies in Coma are given along the top axis of Figure \ref{nuvr}.
Stellar masses were fit with {\tt KCORRECT} \citep{Blanton2007a} using 7-band photometry from {\it GALEX}/SDSS plus
 {\it Spitzer} IRAC 3.6 $\micron$ data when available \citep{Jenkins2007}.
The logarithm of stellar mass is linear with $r$-band magnitude which we fit as Log M$_{*}$ $= 16.364 - 0.435$$\times$$r$ (0.06 dex scatter) for passive galaxies.
The relation for SF galaxies has roughly the same slope but is shifted to lower mass by 0.25 dex with larger scatter (0.2 dex),
e.g., for a given stellar mass, SF galaxies are $\sim$0.5 mag brighter in the $r$-band as compared to passive galaxies.
The majority of confirmed Coma member galaxies have stellar masses larger than M$_{*}=10^{7}$ $\msun$.

\subsection{Missing Dwarf Star Forming Galaxies in Coma?}
Interestingly, both CMDs show a deficit of star forming galaxies in Coma at faint optical magnitudes.
This is especially evident in the region of color space labeled ``gap" where there is a lack of
confirmed SF galaxies.
This region is tilted in the CMDs along a path that traces stellar mass in the range M$_{*}$$\sim$10$^{7.5-8.0}$ $\msun$.
Although selection effects of the spectroscopic redshift coverage may create an artificial void, this seems implausible as 
several confirmed background galaxies are located in this region,
in addition to background galaxies not shown in Figure \ref{nuvr} that were excluded from this study (Section 2.3).
It does not result from a magnitude or surface brightness bias in the redshift coverage of Coma members.
For example, several confirmed Coma members are located at fainter optical magnitudes and
the passive Coma members with stellar masses in the range M$_{*}$$\sim$10$^{7.5-8.0}$ $\msun$ would have
a similar surface brightness to SF galaxies in the gap.

We have performed a {\it post hoc} analysis to estimate the probability that the gap results from statistical chance alone.
For simplicity, we define the gap in the $r$-band interval $18.9<r<19.7$ where there is a
lack of confirmed Coma members (this is a conservative definition as the gap is tilted and thus covers a slightly larger area of the CMD).
The null hypothesis is that the true fraction of confirmed Coma members inside the gap among all galaxies (referred to here as the ``cluster fraction'')
is the average of the cluster fractions on either side.
Restricting our analysis to the region of Figure \ref{nuvr} with colors bluer than NUV$-r=3.5$,
the expected cluster fraction inside the gap is 5\% (7.5\% and 2.5\% for the 0.8 mag bins on either side).
There are 133 total galaxies inside the gap so we expect 6-7 (133$\times0.05$) confirmed Coma members. The probability of zero confirmed Coma members
inside our gap is therefore 0.1\% under binomial statistics.

A more comprehensive Monte Carlo analysis shows that the probability that a similar gap appears at {\it any} r-band magnitude in Figure 5 is 5-10\%.
This worst-case scenario implies that we do not expect the gap at a particular location in the CMD.
There is independent qualitative evidence, however, that a gap should appear near its current location based on the LFs presented in later sections.
For example, the passive galaxy LF in Coma shows a turnover at the same magnitude where the gap extends into the red sequence, and there is suggestive evidence
for a turnover in previous optical LFs in Coma at the gap \citep{Mobasher2003LF}. Deep Virgo UV LFs also show a turnover at magnitudes associated with the gap for both
star forming and passive galaxies \citep{Boselli2011}.  We therefore expect the random chance probability for our gap to be less than the worst-case scenario.

A physical explanation for the gap may be more plausible as suggested by its alignment with the expected evolution of galaxies in a particular stellar mass range.
For example, the gap is aligned with the path that a dwarf star forming galaxy would follow after rapid quenching.
We modeled this track using the GALEV evolutionary synthesis code \citep{Kotulla2009} for a galaxy with constant star formation (12 Gyr)
followed by a 100 Myr quenching event.
The model properties just prior to quenching were selected to match NGC 6822 \citep[M$_{*}=10^{7.66}$ $\msun$, M$_{\mathrm{gas}}=10^{8.24}$ $\msun$;][]{Hunter2004,Hunter2010}, a 
galaxy in the Local Group that has UV-optical colors that place it inside the gap.
The track is shown in Figure \ref{nuvr} and spans 4 Gyr from the onset of quenching (the diagram shows the position of the model after 0, 0.25, 0.5, 1, and 2 Gyr).

Note that it would not be surprising if future redshift surveys identified {\it some} Coma members in this region.
In fact, of the 16 confirmed Coma members at NUV $>21$ in the outer 0\fdg5$-$0\fdg6 {\it GALEX} FOV (thus these galaxies did not satisfy the criteria to be included in this study),
we identified one SF galaxy and it is located inside the gap (albeit just bluer than the SF/passive dividing line).
We therefore consider the gap as representative of an overall decline in the number of dwarf SF galaxies in Coma which is not necessarily exclusive to this region
of color space. The gap is referenced throughout this study primarily as a tool for interpreting features of the UV LFs.  In the Discussion section, we comment further on the possible 
physical origin of the gap and the SF galaxies located at fainter magnitudes.

\subsection{UV$-r$ vs.~UV CMDs}
The UV$-r$ vs.~UV CMDs for both {\it GALEX} bands are shown in Figure \ref{uvcmd}.
These CMDs help to illustrate the procedure invoked in this study for measuring the UV LFs and we will refer back to these diagrams in Section 4.
Here, we make a few brief comments/predictions for the Coma UV LFs.
The gap extends across a large magnitude range in both UV bands which will affect the LFs for star forming galaxies at m$_{\mathrm{UV}}$$\simgt$21, i.e.,
the UV LFs for SF galaxies should have a shallower slope starting at m$_{\mathrm{UV}}\approx21$ (M$_{\mathrm{UV}}=-14$ at the distance of Coma).
We note that there is a high concentration of SF galaxies in Coma near the boundary that separates passive/SF galaxies, e.g.,
most galaxies at m$_{\mathrm{UV}}$$\simgt$21 are located in the upper half of the SF color space (NUV$-r>2.5$ and FUV$-r>3.25$).
We therefore expect a sharp decline in the UV LF where the gap meets this boundary at FUV$\approx24$ (M$_{\mathrm{FUV}}=-11$) and NUV$\approx23$ (M$_{\mathrm{NUV}}=-12$).

Our catalog is progressively more incomplete for blue galaxies at magnitudes fainter than m$_{\mathrm{UV}}=22$ owing to the optical magnitude limit imposed at $r=21.2$ mag.
For measurements of the Coma UV LF, we consider our coverage of SF galaxies in Coma to be complete to slightly fainter UV magnitudes (m$_{\mathrm{UV}}=23$).
This fainter limit excludes only a small wedge of color space not covered by our catalog ($22<m_{\mathrm{UV}}<23$ and colors bluer than m$_{\mathrm{UV}}-r<1.8$).
We do not expect many (if any) Coma members inside this region as very few cluster members have similarly blue colors, and this region corresponds to
starburst galaxies\footnote{We consider starburst galaxies to have specific star formation rates (SSFR) larger than Log [SSFR] $=-9$ yr$^{-1}$.
Coma members in this region of color space would be classified as starburst galaxies based on estimates of the SSFR using the stellar mass relation described in 
\S 3.2 and the \cite{Kennicutt1998} UV SFR conversion assuming negligible dust extinction.} which are quite rare in the region of Coma studied here \citep{Mahajan2010}.
Passive galaxies are the majority population in Coma at m$_{\mathrm{UV}}>23$ so the imposed optical limit at $r=21.2$ has little impact on the LFs for the total cluster population.

\section{Luminosity Function}
Measurements of the cluster LF, defined as the number density of galaxies for a given range of luminosity, require a reliable method of separating cluster members 
and foreground/background galaxies.
Spectroscopic redshifts provide the most direct assessment of cluster membership but are often limited to a subset of the total population.
Past UV LF studies of local clusters have determined membership for the remaining galaxies using spectroscopic completeness corrections \citep[e.g.,][]{Cortese2003b,Cortese2005,Cortese2008b,Boselli2011},
statistical subtraction of field galaxies based on a control field \citep[e.g.,][]{Andreon1999,Cortese2003a}, or selecting cluster member galaxies via morphology \citep{Cortese2008b}.
Morphology selection cannot be used for this study as SDSS lacks the spatial resolution to identify dwarf galaxies at the distance of Coma.
We rely on the spectroscopic completeness method for this study.

We have $\sim$100\% spectroscopic redshift coverage for galaxies brighter than m$_{\mathrm{UV}}=20$ mag. The completeness of the redshift coverage is 
shown in the top panels of Figure \ref{spec_coverage} separated into blue and red galaxies.
Following the spectroscopic completeness method, the number of cluster members at fainter magnitudes is estimated statistically by assuming that the
fraction of confirmed Coma members among galaxies with redshifts is representative of the entire photometric catalog
\citep{DePropris2003,Mobasher2003LF}. Specifically, we scale the photometric catalog by the membership fraction defined as:
\begin{equation}
f(m)= \frac{N_{c}(m)}{N_{s}(m)}_{,}
\label{mf}
\end{equation}
where N$_{c}(m)$ is the number of confirmed Coma member galaxies, and N$_{s}(m)$ is the number of galaxies with spectroscopic redshifts.
The membership fractions are shown in the bottom panels of Figure \ref{spec_coverage}.

The luminosity function is the number of galaxies in the  {\it GALEX}/SDSS photometric catalog (N$_{p}$) scaled by the membership fraction, UV coverage area (A),
and the magnitude bin size ($\Delta$$m$=0.5 mag):
\begin{equation}
\Phi(m)= f(m)~\frac{N_{p}(m)}{A ~ \Delta m}
\label{lf_eqn}
\end{equation}
\begin{equation}
\frac{\delta\Phi(m)}{ \Phi(m)} = \left[\frac{1}{N_{p}(m)} + \frac{1}{N_{c}(m)} - \frac{1}{N_{s}(m)}\right]^{1/2}_{,}
\label{lferr}
\end{equation}
where the error formula assumes N$_{p}$ is a Poisson variable and N$_{c}$ is a binomial variable (i.e.,~the number of successes in N$_{s}$ trials with probability $f$).

\begin{figure}[t!]
\vspace{30pt}
\centering
\includegraphics[angle=0.,scale=0.75]{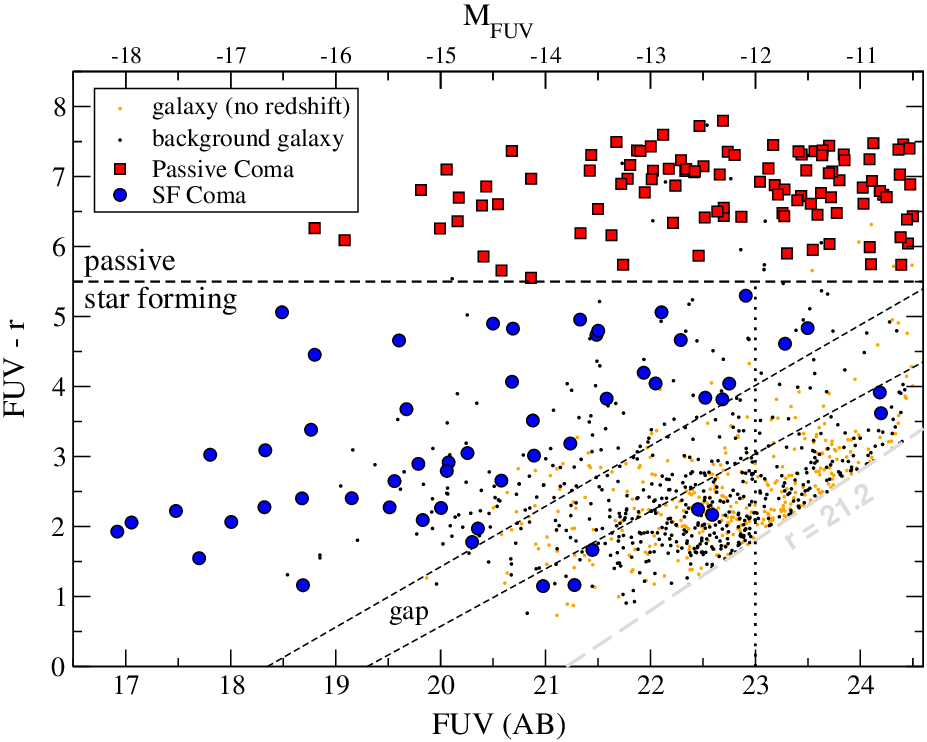}
\\[20pt]
\includegraphics[angle=0.,scale=0.75]{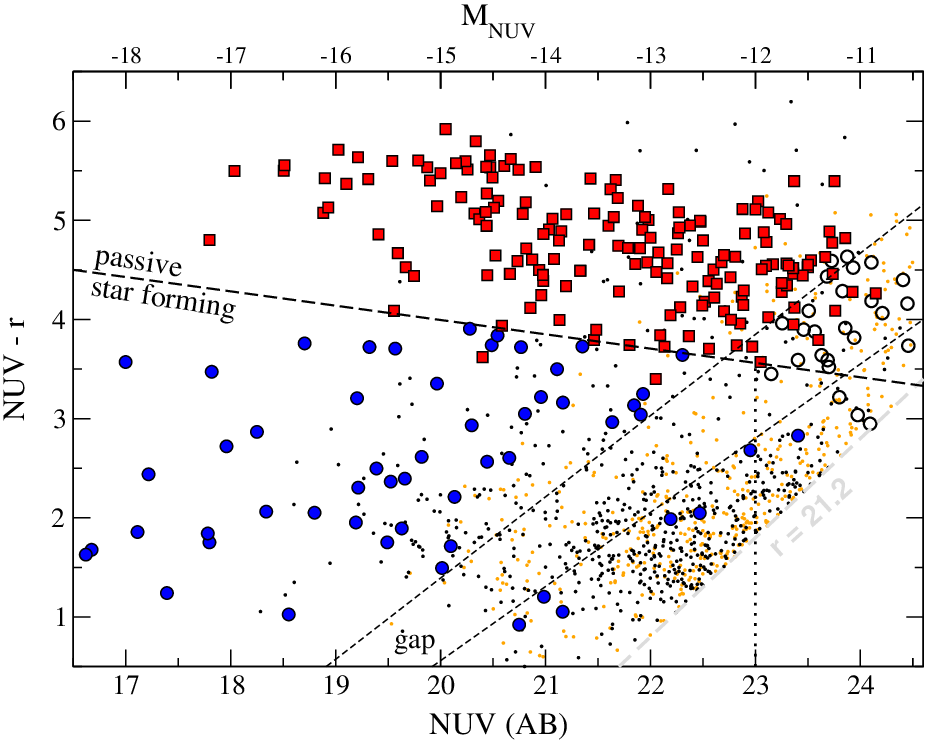}
\caption{\label{uvcmd} UV-optical CMDs plotted against {\it GALEX} FUV (left panel) and NUV (right panel) magnitude.
Symbols are identical to Figures \ref{fuvr} and \ref{nuvr}. The top x-axis gives the corresponding absolute magnitudes
for galaxies at the distance of Coma.
These plots are useful for illustrating the color distribution of galaxies in UV intervals and the method invoked for measuring the UV LFs in Section 4.
The gray long-dashed lines show the SDSS $r$-band magnitude limit ($r$=21.2) adopted for this study.
The black horizontal long-dashed lines are used to separate passive/SF galaxies for the UV LF
measurements, and were selected to closely match the FUV$-r$ color classification of Coma members as shown on the diagrams.
The vertical black dotted lines indicate the completeness limit adopted for SF galaxies; at fainter magnitudes, our
galaxy sample is progressively more incomplete owing to the $r$-band magnitude limit at $r=21.2$.
The black short-dashed lines enclose the color space affected by the gap which results in fewer SF Coma members at 
magnitudes fainter than M$_{\mathrm{UV}}\simgt-14$.
}
\end{figure}

\begin{figure}[t!]
\vspace{50pt}
\centerline{\includegraphics[angle=0.,scale=1]{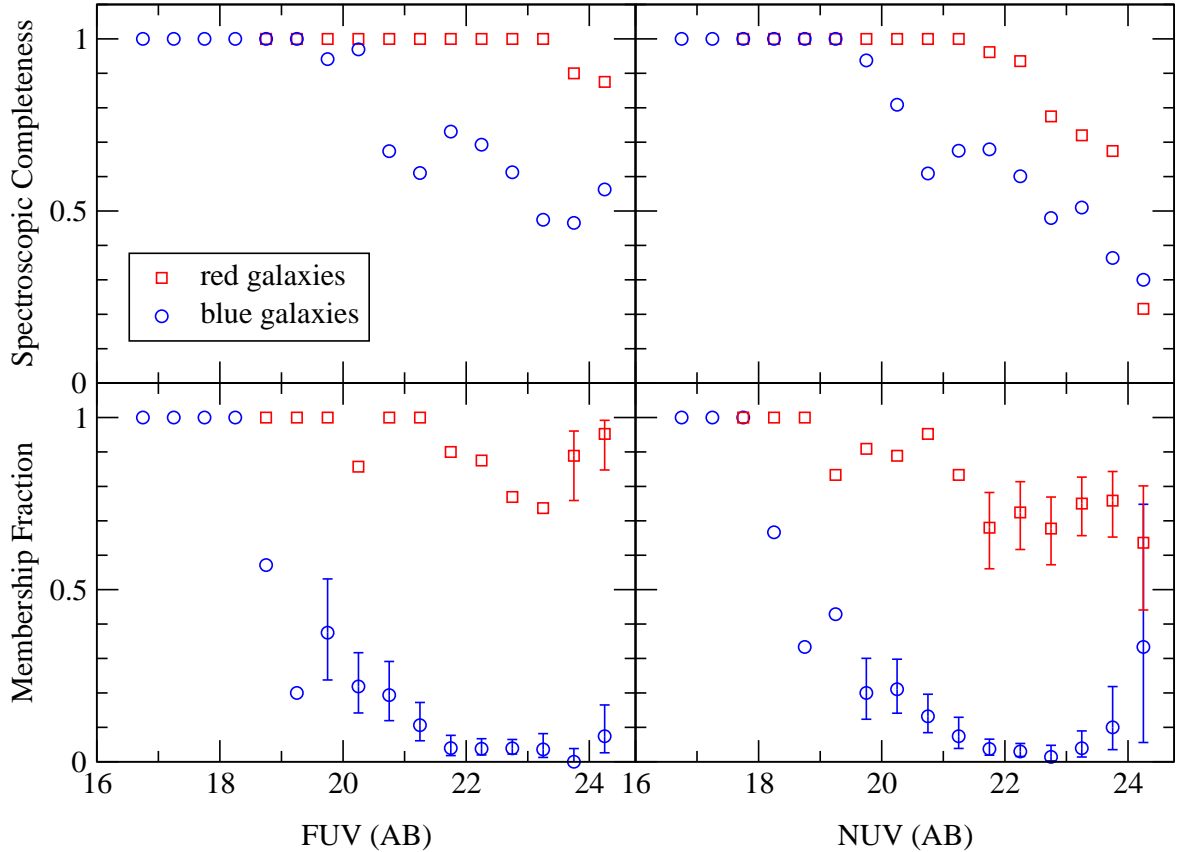}}
\caption{\label{spec_coverage} Properties of the spectroscopic redshift coverage at FUV (left panels) and NUV (right panels) wavelengths.
The spectroscopic completeness - the fraction of all galaxies with spectroscopic redshifts - is shown in the top panels ($N_{s}/N_{p}$).
The membership fraction - the fraction of galaxies with redshifts that are cluster members - is shown in the bottom panels ($N_{c}/N_{s}$).
Measurements are performed separately for blue/red galaxies using the color separation defined in Figure \ref{uvcmd}.
The uncertainties are binomial errors from the formalism in \cite{Gehrels1986};
error bars are shown only at magnitudes where we use membership fractions to measure the LF (i.e., where the spectroscopic completeness is less than 100\%).
The top panels indicate that red galaxies have higher spectroscopic completeness as compared to blue galaxies, and bottom panels show that a
significantly higher fraction of red galaxies are Coma members.
Given the color dependence of the spectroscopic redshift coverage, we construct the Coma UV LF as a summation of the separate blue/red LFs
in order to avoid systematic color selection effects.}
\end{figure}

We must also account for the systematic bias related to measuring the cluster LF in a different band than used for redshift target selection (the optical $r$-band).
Specifically, the redshift coverage at faint UV magnitudes is better for optically-bright red galaxies (primarily cluster members) 
which would overestimate the membership fraction for blue galaxies that are primarily background galaxies.
We note that this color selection bias would be enhanced by including galaxies fainter than the spectroscopic
coverage, i.e., this would simply add more blue galaxies without redshifts.
We account for the color bias by measuring the LF separately for star forming and passive galaxies using the dividing lines shown in Figure \ref{uvcmd} to separate galaxies by UV$-r$ color.
The total LF is a summation of the color-dependent LFs:

\begin{equation}
\Phi(m)=\sum_{m-r}^{sf,p}\frac{f(m | m-r)~N_{p}(m | m-r)}{A~\Delta m}.
\label{lfcoloreqn}
\end{equation}
This formula reduces to eqn.~(2) when the membership fraction is independent of UV-optical color.
A significant difference for the LFs measured using eqn.~(2) and eqn.~(4), as was identified in this study,
indicates bias owing to color selection effects and the color-dependent equation should be adopted.

\begin{figure}[t!]
\vspace{30pt}
\centerline{\scalebox{0.72}{\includegraphics*[6,16][480,374]{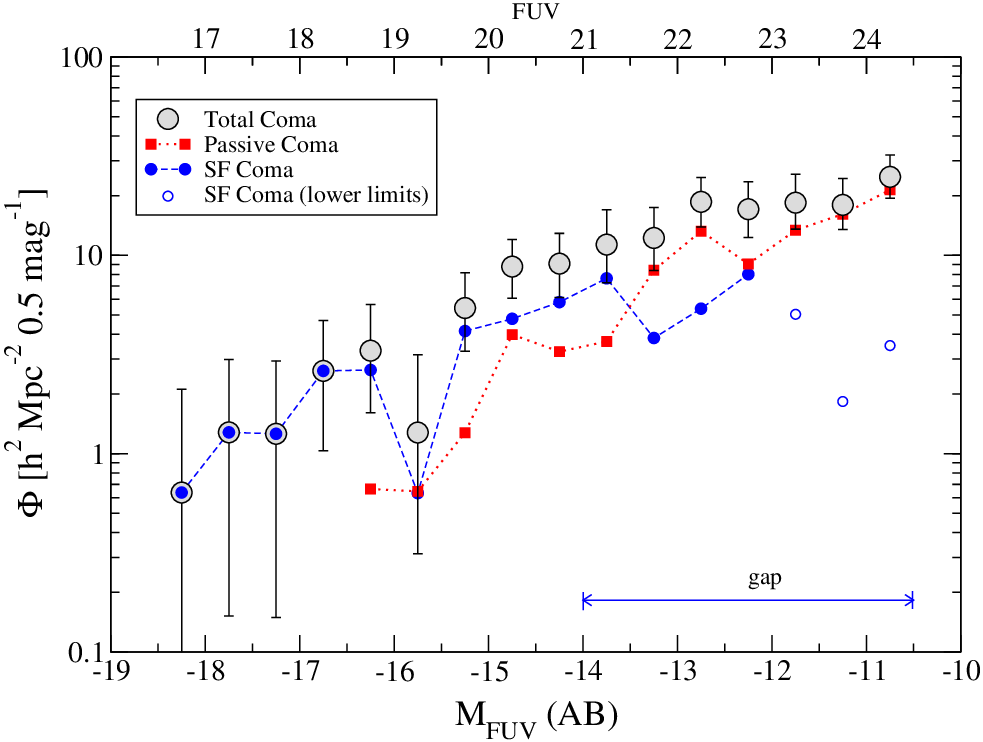}}}
\vspace{15pt}
\centerline{\scalebox{0.72}{\includegraphics*[6,16][480,374]{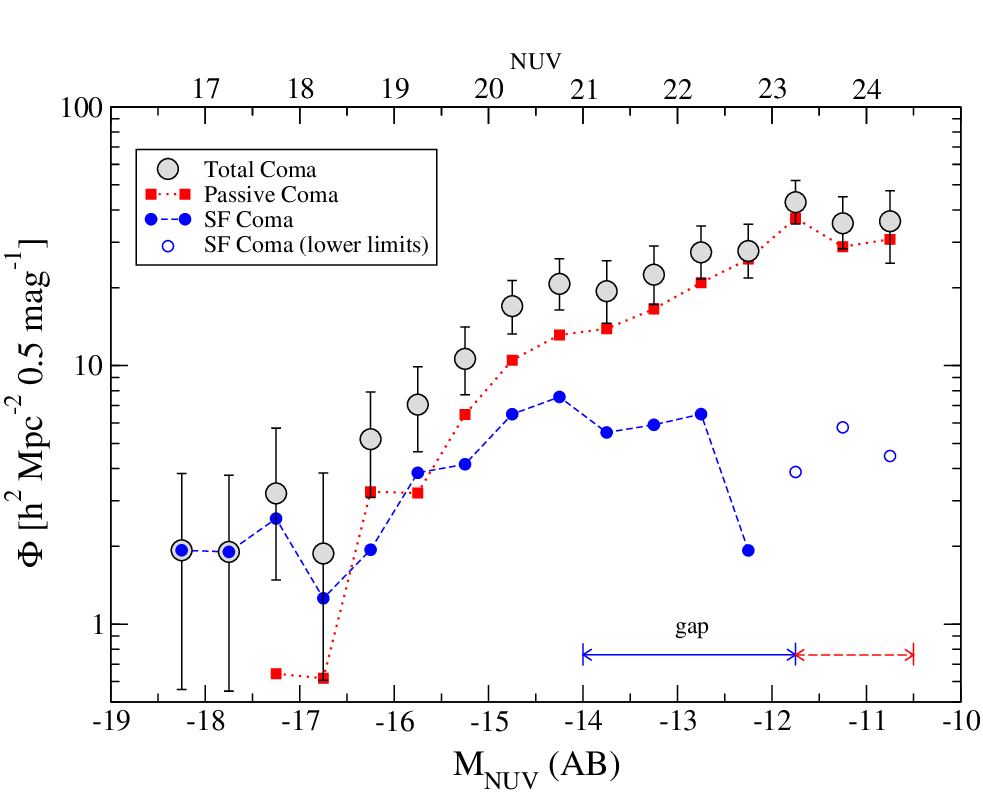}}}
\caption{\label{lf} LFs for the Coma cluster (large gray-filled circles) in the {\it GALEX} FUV (left) and NUV (right) bands.
We also show the separate LFs for passive galaxies (red filled squares) and SF galaxies (blue filled circles); open blue circles
are lower limits owing to incomplete coverage of SF galaxies at M$_{\mathrm{UV}}>-12$.
The magnitude intervals that are affected by the gap are indicated at the bottom of each panel; the SF LFs show a turnover 
at magnitudes associated with the gap. The NUV interval where the gap extends into the red sequence (see Figure \ref{uvcmd})
is shown as a dashed line and corresponds to a drop in the passive NUV LF (the gap does not extend into the red sequence in the FUV band).}
\end{figure}

In Figure \ref{lf}, we show the UV LFs for Coma in both {\it GALEX} bands, including the separate measurements for
passive and star forming galaxies.
The LFs have been corrected for the Eddington Bias with errors conservatively taken as 50\% of the correction \citep[][]{Eddington1913},
and also for imperfect SDSS and {\it GALEX} detection efficiency as described in the Appendix.
The corrections amount to less than 7\% across the full magnitude range.
Separate symbols are used for SF galaxies at magnitudes fainter than M$_{\mathrm{UV}}\approx-12$
where our catalog has incomplete coverage (see Section 3.4). The LF values are given in Table 1.

The total LF in each band shows a slight dip at bright magnitudes (M$_{\mathrm{FUV}}\sim-15.7$ and M$_{\mathrm{NUV}}\sim-16.7$).
The galaxies brighter than this feature consist almost entirely of SF galaxies.
A similar Gaussian-like distribution of bright spirals in Coma has been observed in previous UV LFs \citep[e.g.,][]{Andreon1999,Cortese2008b},
and in the Coma LF at radio wavelengths \citep{Miller2009b}.
The SF LFs increase again at fainter magnitudes before another turnover at M$_{\mathrm{UV}}$$\simgt$$-14$.
This faint turnover and the gap start at roughly the same magnitude as predicted in Section 3.4.
The magnitude range associated with the gap is labeled on the diagrams.
There is an abrupt drop in the SF LFs at fainter magnitudes (M$_{\mathrm{FUV}}\approx-11$ and M$_{\mathrm{NUV}}\approx-12$) 
which was also predicted in Section 3.4; these magnitudes correspond to the location in the UV CMDs (Figure \ref{uvcmd}) where the gap meets the
boundary that separates SF/passive galaxies, where there is otherwise a more prevalent number of SF galaxies in Coma.
Interestingly, passive galaxies in the NUV band flatten at magnitudes where the gap extends into the red sequence (M$_{\mathrm{NUV}}\simgt-11.5$).
Our interpretation of this feature is given in the Discussion section.

\begin{figure}[t!]
\vspace{50pt}
\centerline{\scalebox{0.65}{\includegraphics*[6,16][480,374]{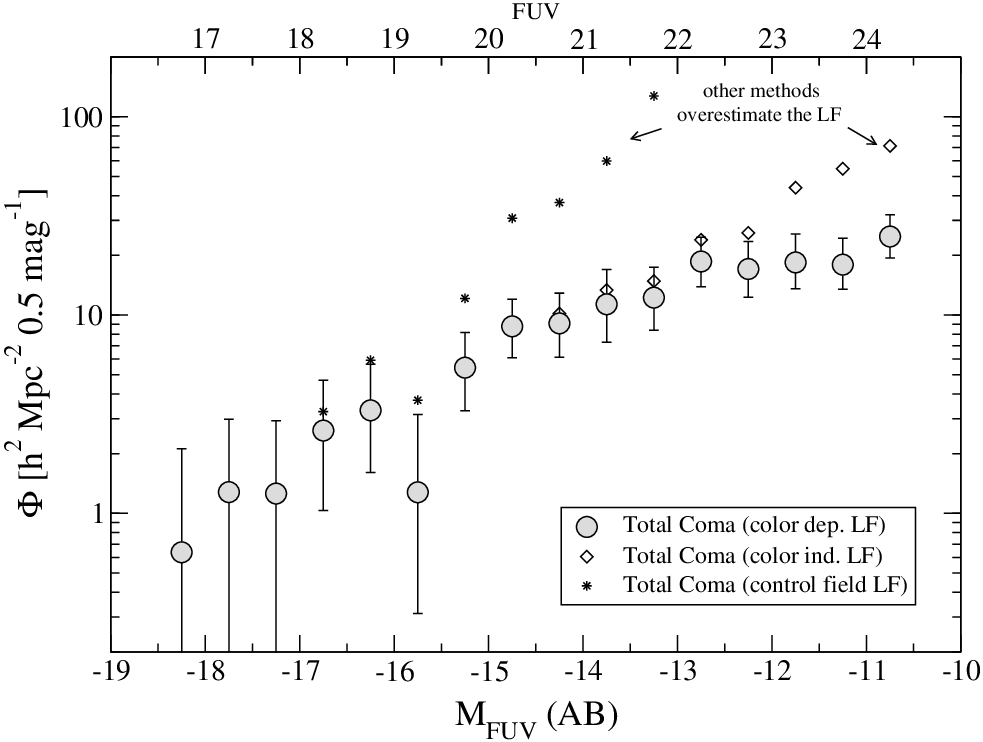}}}
\caption{\label{otherlf} Comparison of Coma FUV LFs that were measured using three different techniques.
Large gray-filled circles show the FUV LF measured using the color-dependent spectroscopic completeness method (the method adopted for this study).
The two alternative LFs are measured using the color-independent completeness method (diamonds) and the subtraction
of a control field (asterisks). The alternative methods clearly overestimate the LF and result in steeper faint-end slopes.}
\end{figure}

We demonstrate the importance of choosing the correct method for measuring the LF in Figure \ref{otherlf}.
This diagram compares Coma FUV LFs that were measured using the color-dependent spectroscopic completeness method
adopted for this study (eqn.~\ref{lfcoloreqn}), and also using both control field subtraction and the color-independent spectroscopic
completeness method (eqn.~\ref{lf_eqn}).
For the control field LF, we subtracted the {\it GALEX} field counts of \cite{Xu2005}
from the galaxy counts in our FOV (Paper I).
This method clearly overestimates the number density and faint-end slope of the LF
probably due to large-scale structure behind Coma.
The color-independent completeness method also overestimates the UV LF as it does not account for 
systematic errors from selecting redshift targets in a different band.
For this method, we also included galaxies to $r$=22.2 (i.e., 1 mag fainter than the redshift coverage)
to test the impact on the LF. We found that the overestimate across the last two magnitude bins is almost entirely
due to including galaxies fainter than the redshift coverage. LF studies must be careful to avoid the systematic effects associated with these other methods.


\subsection{LF Fitting}
We have modeled the Coma UV LFs as both a single Schechter function and a two-component distribution
consisting of a Gaussian at bright magnitudes and a Schechter function at faint magnitudes.
In Figure \ref{uvlf_fit}, we show the fits to the total, star forming, and passive LFs in the FUV and NUV bands.
Fits were performed using the IDL package {\tt mpfit} which relies on the Levenberg-Marquardt technique to solve the least-squares problem \citep{Markwardt2009}.
The fitted parameters for each model are listed in Table 2.  The details of the fitting procedures are given below.

\subsubsection{Single Component Schechter Functions}
We modeled the LFs as a single Schechter function \citep{Schechter1976} in the following parametric form:
\begin{equation}
\Phi(M)=\phi^{*} X^{\alpha+1} e^{-X},
\label{lumfunc}
\end{equation}
where $X=10^{-0.4(M-M^{*})}$, $\phi^{*}$ is the normalization, $M^{*}$ describes the bright-end turnover, and $\alpha$ is the faint-end slope.

For the total UV LFs, our data do not adequately constrain the bright-end turnover owing to the limited
spatial coverage of the cluster. We therefore fixed M$^{*}$ to
values reported in a wide-field study of Coma \citep[M$^{*}_{\mathrm{NUV}}$ $=-18.5$ and M$^{*}_{\mathrm{FUV}}$ $=-18.2$;][]{Cortese2008b}
and solved for the faint-end slope. This same procedure was applied to the star forming LFs.
For the passive galaxy population, the bright-end turnover is well constrained which allowed us to solve simultaneously for the Schechter parameters.

\subsubsection{Two Component Gaussian + Schechter Functions}
LFs are known to be a superposition of Gaussian functions for normal galaxy types (E-S0-Sa-Sb-Sc-Sd)
and Schechter functions for dwarf galaxies \citep[dEs and Irr-BCDs;][]{Binggeli1988,DeLapparent2003}. The relative abundances
of these galaxy types vary with environment resulting in the different shapes of field and cluster LFs. We have therefore fit the UV
LFs with a two-component model that consists of a Gaussian function at bright magnitudes and a Schechter function at faint
magnitudes. For the SF and passive LFs, the Schechter component is aligned with the dwarf galaxy population in order to match the 
expected distribution of these galaxy types. This provides a more physical interpretation of the fitted parameters.
The bright Gaussian component is relatively less constrained owing to small number statistics but such galaxies are not the focus of this study.
The total UV LFs include a mix of normal and dwarf galaxies at faint magnitudes so the fits do not model individual galaxy types.
The two-component model does, however, provide a better parametrization of the shape of the total UV LF, e.g., the bright spiral population
is clearly distinct from the other galaxies as noted in previous UV LF studies \citep[e.g.,][]{Cortese2008b},
and is now well matched by the two-component model as seen in Figure \ref{uvlf_fit}.  A more detailed description of the fitting  procedure
is given below.

For the total and SF LFs, we confined the center of the Gaussian to magnitudes brighter than M$_{\mathrm{UV}}=-16$, but otherwise fit the parameters without constraints.
A skew term was added to the Gaussian function for the FUV LF to match the observed distribution.
The last magnitude bin of the NUV SF LF drops abruptly so we performed fits both with and without this data point.
The faint-end slope is $\alpha=-0.48$ and --0.72 for the full magnitude range and truncated range, respectively.
The preferred version is a matter of function as the fit across the full magnitude range is a better description of our data to the completeness limit, whereas 
the truncated version is more consistent with the lower limits given at fainter magnitudes (the truncated fit is shown in the diagram).
Both fits are given in Table 2.

The NUV LF for passive galaxies shows an inflection near M$_{\mathrm{NUV}}$ $=-14$.
This location corresponds to the magnitude where dwarf galaxies become the dominant population.
We therefore fit the normal passive galaxy population with a Gaussian (M$_{r}<-18.5$; shown as open diamonds in Figure \ref{uvlf_fit}), and then
fit the remaining dwarf galaxies with a Schechter function.
In contrast to the total and SF LFs, the shape of the passive NUV LF is matched equally well using the single or two-component model.
We were unable to perform a two-component fit to the passive FUV LF as dwarf galaxies are the majority in
only the last few magnitude intervals.


\subsection{LF Discussion}
The most notable difference between the single and two-component models are the much shallower 
faint-end slopes predicted by the Gauss-plus-Schechter functions for all Coma UV LFs.
The two-component model removes the sensitivity of the Schechter power law to bright spiral galaxies which tend
to force steeper faint-end slopes in order to match the whole LF. This model gives faint-end
slopes of $\alpha$ $\approx$ --1.15 in both {\it GALEX} bands for the total UV LFs, as opposed to $\alpha$ $=-1.39$
for the single Schechter model. The two-component fits for the separate SF and passive galaxy
populations both give faint-end slopes shallower than $\alpha$ $=-1$ (with $\sim$83\% confidence for the star forming population).

A similar faint-end slope was reported for dwarf SF galaxies in an H$\alpha$ LF study at the Coma center \citep[$\alpha$$=-0.60$;][]{Iglesias2002}.
This turnover of the H$\alpha$ LF supported the notion that dwarf galaxies were tidally destroyed near the center of Coma resulting in a radial gradient of the faint-end slope
\citep[e.g.,][]{Thompson1993}.
That we have measured a similar faint-end slope at the infall region suggests that a radial gradient for dwarf SF galaxies is either smaller than expected or does not exist.
A deep {\it GALEX} observation was recently performed at the center of Coma (where we have similar redshift coverage) which will allow us to
test the radial dependence of the faint-end slope for dwarf galaxies that span a large range of cluster-centric distance.
We may further constrain this notion by re-measuring the Coma LFs in bands
that were previously associated with a radial gradient of the faint-end slope \citep[e.g., $U$-band, 3.6$\micron$, 24$\micron$;][]{Beijersbergen2002,Jenkins2007,Bai2006},
but using the more reliable measurement techniques presented here and the improved redshift coverage that is now available.

We are not aware of a previous Coma LF study that reported a faint-end slope shallower than $\alpha=-1$ for dwarf passive galaxies.
However, this turnover is obvious in several previous optical LFs in Coma \citep[see the $R$-band LF compilation in Figure 8 of][]{Milne2007}.
It was not reported as most studies extended the Coma LF to much fainter magnitudes via the (unreliable) method of control field subtraction,
which resulted in a sharp upturn of the LF after the turnover.
\cite{Trentham1998} showed that the Coma LF is relatively flat after the turnover when considering low surface brightness (LSB) galaxies alone
(a more reliable tracer of cluster membership; see their Figure 4d). This is more consistent with the Coma NUV LFs presented in this study.

In the next section, we compare our results to previous UV LF studies.
We note that comparing parametrized fits are difficult owing to their dependence on magnitude, and hence are
reliable only when considering the same magnitude limit (or similar galaxy types for LFs in different bands).
Moreover, although the two-component models provide a better description of the Coma UV LFs,
most previous studies have used a single Schechter model.
For these reasons, we also rely on comparisons using the actual LF data points.


\subsubsection{Comparisons to Other UV LFs}
In Figure \ref{uvlf_other}, we compare our Coma UV LFs to previous {\it GALEX} LFs measured for local galaxy clusters.
The other LFs are the wide-field study of Coma \citep[r$\sim$0.1-1.5 R$_{\mathrm{vir}}$;][]{Cortese2008b},
and studies at the centers of the relatively young Abell 1367 cluster \citep[r$\sim$0.0--0.7 R$_{\mathrm{vir}}$; $\bar{\mathrm{r}}=0.3$ R$_{\mathrm{vir}}$;][]{Cortese2005}
and Virgo cluster \citep[$\bar{\mathrm{r}}=0.3$ R$_{\mathrm{vir}}$;][]{Boselli2011}.
The UV LFs are similar at bright magnitudes despite both the different richness of the clusters and radial coverage
of our study (r$\sim$0.2--0.9 R$_{\mathrm{vir}}$; $\bar{\mathrm{r}}=0.55$ R$_{\mathrm{vir}}$).
For example, the faint-end slopes of our Coma LFs are nearly identical to the wide-field Coma study \citep[$\alpha_{\mathrm{NUV}}=-1.77$ and $\alpha_{\mathrm{FUV}}=-1.61$;][]{Cortese2008b}
and the A1367 study \citep[$\alpha_{\mathrm{NUV}}=-1.64$ and $\alpha_{\mathrm{FUV}}=-1.56$;][]{Cortese2005} when
we adopt the same magnitude limits at M$_{\mathrm{UV}} = -14$ and --13.5, respectively. We refit our Coma UV LFs
at these limits and recovered faint-end slopes of $\alpha \simeq -1.70$ and --1.60 in both {\it GALEX} bands, respectively.
The Coma LFs flatten at fainter magnitudes resulting in the shallower slopes that we report across our full magnitude range.
The Virgo LFs reach similar magnitude limits as our study but are relatively flat, probably owing to fewer dwarf passive galaxies
as reported in a recent optical LF study of local galaxy clusters \citep{Weinmann2011}.

It is easier to assess the LFs at faint magnitudes by separating the star forming and passive galaxy populations.
In Figure \ref{nuvlfpts} (left panel), we compare the NUV LFs for star forming galaxies in Coma to measurements performed for Virgo and the local field \citep[$z$$<$0.1;][]{Treyer2005}.
We derived another NUV LF for the local field environment based on the SDSS $r$-band field LFs \citep[$z$$\simlt$0.05;][]{Blanton2005}.
The optical LFs were converted to the NUV band using the NUV$-r$ color LFs in \citet[][]{Wyder2007}. The shaded region in Figure \ref{nuvlfpts}
represents the limits of the derived field UV LF with the bottom/top edges corresponding to the raw and completeness corrected $r$-band LFs,
respectively \citep[i.e., ``LF1 blue" for the bottom edge and ``LF3 total" at M$_{r}$$>$-18 minus 20\% to remove red galaxies for the top edge;][]{Blanton2005}.
The field LFs are similar despite the different selection criteria and the local volumes that were sampled.

Interestingly, the LFs of Coma and Virgo have similar shapes and show the same turnover at M$_{\mathrm{NUV}}$$=-14$.
We did not normalize the Virgo LF in Figure \ref{nuvlfpts} thus both clusters have nearly the same number density of SF galaxies.
\cite{Cortese2008b} suggested that the SF LFs of local clusters should be identical as they are tracing the same field population at the moment of accretion prior to rapid quenching,
i.e., we detect SF galaxies in a shell surrounding the center of each cluster. This implies that although the Virgo study
covered the center of the cluster, the SF galaxies for both clusters are detected in projection and are not located inside the core region.
Assuming this is correct, we can make the following points.
The Virgo LFs were separated by morphology, as opposed to SSFR for Coma, which suggests that the timescales for morphological transition and quenching are roughly similar.
We also note that there is mild agreement between the cluster LFs at faint magnitudes (M$_{\mathrm{NUV}}$$>$--12) despite incomplete coverage of Coma SF galaxies.
The incomplete coverage applies primarily to dwarf SF galaxies with relatively high SSFRs which suggests that such galaxies are rare in the region of Coma studied here,
i.e., the lower limits given for Coma SF galaxies at M$_{\mathrm{NUV}}$$>-12$ are probably near the true value.

The field LFs trace both clusters at magnitudes brighter than M$_{\mathrm{NUV}}=-14$ but increase across magnitudes associated with the cluster turnover.
{\it This is direct evidence that dwarf SF galaxies are more susceptible to environmental processes than normal galaxies.}
The divergence of the cluster and field LFs appears to contradict the results of \cite{Cortese2008b} who suggested they should be identical (albeit this relation was 
established at magnitudes brighter than the turnover).
Two possible explanations for this discrepancy are:
(1) the average field LF is not representative of the infalling population onto local galaxy clusters at faint magnitudes (i.e., a turnover exists prior to reaching the infall region studied here),
or (2) the average field LF describes the infall population but the quenching rate exceeds the accretion rate for galaxies associated with the turnover
(as opposed to higher-mass galaxies that may be in equilibrium).
If the second scenario is correct and the number of infalling galaxies is conserved (i.e., blue galaxies are quenched but not destroyed),
then we expect the LF for passive galaxies to follow the field LF and increase across magnitudes associated with the turnover for SF galaxies.

The right panel of Figure \ref{nuvlfpts} shows the same diagram but now includes dwarf passive galaxies for both Coma and Virgo.
For an easier comparison, we have shifted the passive LFs brighter by 2.5 mag which is the average color difference for dwarf passive/SF galaxies at the turnover
(this shift roughly aligns the LFs by stellar mass).
The shaded region is the expected distribution of dwarf passive galaxies in the cluster for a quenched infall population that
follows the field UV LF (assuming blue galaxies are quenched but not destroyed).
This model consists of the blue field LF shown in the left panel plus field galaxies that are passive before entering the cluster (shifted by 2.5 mag);
the model LF was then normalized to the Coma UV LF for passive galaxies at magnitudes brighter than the turnover.
The diagram shows that the passive LFs for both clusters do not trace the quenched field model at faint magnitudes.
We conclude that the infall population onto local clusters is not well described by the average field LF inside the virial radius.

The more likely scenario is that the infall population shows a turnover prior to reaching the virial radius. Hence, the passive LF in Coma
should roughly follow the same distribution as dwarf SF galaxies.
From Figure \ref{nuvlfpts}, the LFs for both populations in Coma have similar shapes and both show the same turnover (after shifting the passive LF by 2.5 mag).
The red fraction -- the fraction of Coma member galaxies that are passive -- varies between only 80-84\% for dwarf galaxies measured across 3 magnitudes
in NUV ($-16<$ M$_{\mathrm{NUV}}<-13$ in Figure \ref{nuvlfpts}).
We note that the Virgo UV LFs for both galaxy types also follow the same distribution before and after the turnover.


\section{Discussion}
Perhaps the most interesting result of this study is the deficit of dwarf SF galaxies in Coma at stellar masses below M$_{*}$$\sim$10$^{8.0}$ $\msun$ as compared to the field environment.
This deficit was identified via UV LFs which show a turnover for dwarf SF galaxies in
Coma at M$_{\mathrm{UV}}\approx-14$ as compared to the rising field UV LFs (Figure \ref{nuvlfpts}).
There is suggestive evidence for a turnover in previous LF studies of Coma.
For example, optical LFs for color-selected blue galaxies in Coma show a turnover at faint magnitudes \citep[M$_{\mathrm{R}}>-16$; see Figure 9 in][]{Mobasher2003LF}.
Although the optical turnover consists of only one data point before reaching the magnitude limit of the \cite{Mobasher2003LF} study,
it was identified for galaxies at both the center of Coma and the infall region studied here.
\cite{Iglesias2002} also reported a turnover for a deep H$\alpha$ LF study at the center of Coma.
The deep UV LFs measured for the Virgo cluster \citep{Boselli2011} are remarkably similar to the Coma LFs presented here and show
the same turnover at M$_{\mathrm{UV}}\approx-14$.
A deficit of dwarf star forming galaxies may be a common feature of local galaxy clusters.

Interestingly, the shapes of the separate UV LFs for dwarf passive and dwarf SF galaxies in Coma are similar and both populations 
show the same turnover after correcting for color differences, e.g., the red fraction of dwarf galaxies is roughly constant (80--84\%) across
3 magnitudes in NUV covering the magnitude range both before and after the turnover.
The UV LFs for both dwarf populations in the Virgo cluster also share the same turnover \citep{Boselli2011}.
This suggests that the dwarf subtypes share a common origin, and that recent quenching of infalling dwarf SF galaxies
is sufficient to build the dE population at the infall region of Coma.
Studies have offered alternative formation mechanisms for dE galaxies that are not connected to the current infall population.
For example, some dE galaxies may have: joined the red sequence at much earlier epochs and are now leaving the core region on a trajectory away from the cluster center
\citep[``backsplash galaxies'';][]{Pimbblet2011}, formed from material that was ejected from massive galaxies via tidal interactions \citep[``tidal dwarf galaxies''; e.g.,][]{Duc2007},
or were members of the proto-cluster that were quenched by photoionization at the epoch of reionization \citep[``squelched galaxies"; e.g.,][]{Tully2002}.
These alternative processes are not consistent with the matched LFs for dwarf SF and dwarf passive galaxies. We conclude that these other processes 
are not important at the cluster-centric distances studied here, but could potentially be relevant in the dense Coma core.

{\bf What are the expected properties of dwarf SF galaxies that are missing in Coma?}
The turnover of the Coma UV LFs is associated with a void, or gap, of dwarf SF galaxies in the UV-optical CMDs (Figures \ref{fuvr} and \ref{nuvr}).
The gap is aligned roughly perpendicular to the red sequence in a small magnitude range at --16.5$\simlt$ M$_{r}\simlt-15.0$ which corresponds to stellar masses
 of M$_{*}$$\sim$10$^{7.5-8.0}$ $\msun$.
The gap itself accounts for only a fraction of the total missing dwarf SF galaxies in Coma. We estimate that $\sim$50 dwarf SF galaxies are required to account for the difference 
between the Coma and field LFs at $-14<$ M$_{\mathrm{NUV}}<-12$, or $\sim$70\% of all SF galaxies in Coma brighter than M$_{\mathrm{NUV}}=-12$.
However, the turnover of the UV LF for SF galaxies begins at magnitudes associated with the gap, which suggests these two features are related.
This allows us to assess the identity of the missing dwarf SF population via comparisons to field galaxy studies.

In Figure \ref{GAP}, we compare the UV-optical CMD of Coma member galaxies to a local sample
of HI-selected dwarf SF galaxies \citep[Sm-BCD-Im/Irr;][]{Hunter2010}.
The Sm field galaxies tend to be located at brighter magnitudes than the gap, whereas Im/Irr galaxies (hereafter dIrr galaxies)
are the majority of galaxies inside or fainter than the gap with a smaller fraction of BCDs.
The dIrr galaxy population is therefore the likely candidate for missing dwarf SF galaxies in Coma.
We are unable to characterize the morphology of most Coma members inside or fainter than the gap, however, as these galaxies are only marginally resolved by SDSS.
At least one galaxy is a probable BCD based on its strong H$\alpha$ emission (EW [H$\alpha$] $\sim500$ \AA; labeled on Figure \ref{GAP}) and high surface brightness relative to other Coma members in this region.
Interestingly, a large fraction of BCDs in the Virgo cluster are located in the same region of color space as our BCD candidate, and a few Virgo 
galaxies classified as ``dE galaxies with blue centers" would be located inside our gap \citep{LiskerIII,Kim2010}.
This may indicate that the gap represents a deficit of low surface brightness SF galaxies in Coma, and those SF galaxies inside or fainter than this region
tend to be more compact galaxies.

{\bf What processes are responsible for the lack of dwarf SF galaxies in Coma?}
Dynamical processes inside the cluster provide an obvious method for removing dwarf SF galaxies.
ICM processes (not tidal) are the dominant mechanism at the off-center region of Coma studied here \citep[e.g., ram-pressure stripping, viscous stripping, thermal evaporation;][]{Boselli2006a}.
ICM processes quench star formation by removing the gas supply to galaxies resulting in a transition to the red sequence.
The thermal evaporation of gas via the hot ICM \citep{Cowie1977} may be more efficient in Coma relative to other local clusters
owing to the high temperature of its ICM \citep[e.g.,][]{Boselli2006a}.
This process is tempered by galactic magnetic fields thus dIrr galaxies are especially vulnerable owing to their low field strengths \citep[a factor of $\sim$3 lower than spirals;][]{Chyzy2011}.
Preventing the infall of external gas may also effectively suppress star formation in dwarf galaxies, possibly
via thermal evaporation of the extended HI gas reservoirs that envelope dIrr and BCD galaxies \citep[e.g.,][]{Begum2008}.
The relative lack of BCD galaxies in Coma as compared to Virgo is consistent with this scenario.
Regardless of the quenching process, gas ablation of infalling galaxies is consistent with the similar shapes of the LFs for dwarf passive and dwarf SF galaxies in Coma.
It does not explain, however, the relative deficit of dwarf SF galaxies in Coma as compared to the field.

Alternatively, the infalling galaxy population onto Coma may have a deficit of dwarf galaxies prior to accretion, possibly
owing to ``pre-processing" of galaxies that flow along filaments feeding the cluster.
Massive clusters such as Coma tend to be associated with larger numbers of intercluster filaments \citep[e.g.,][]{Pimbblet2004}.
Filaments may in fact connect the Coma cluster, embedded in the Great Wall along with the A1367 cluster \citep{Geller1989}, 
to a large structure of galaxies at $z\sim0.05$ and the SDSS Great Wall at $z\sim0.08$ \citep{Gott2005} and beyond \citep{Adami2009}.
The infall region of Coma studied here also lies in projection with an intercluster filament that connects Coma to the A1367 cluster, and 
may be specifically affected by a subcluster merging process associated with the NGC 4839 group along this filament.

Galaxies located in intercluster filaments show increased SF activity at the periphery
of clusters ($\sim$1.5-2 R$_{\mathrm{vir}}$) possibly due to galaxy harassment with infalling substructure \citep{Porter2008}.
The periphery of the Coma cluster, for example, hosts a large fraction of post-starburst (k+A) galaxies as compared to both the surrounding field and
the intercluster filament that connects Coma and A1367 \citep{Mahajan2010,Gavazzi2010}.
After this initial spike in star formation, harassment can rapidly remove up to 50-90\% of the stellar mass in dwarf galaxies \citep{Moore1999}.
The remnants of dIrr galaxies associated with this process (if they survive complete disruption) would lie on the red sequence at fainter UV magnitudes than detected in our study.
These remnants may be the amorphous very LSB galaxies (VLSBs) that have been detected in both Virgo and Coma at optical wavelengths
 \citep{Sabatini2005, Hammer2010b}.
Preliminary optical LFs from the HST-ACS Coma Treasury survey (Trentham et al.~2011, in preparation) show an increasing number of VLSB galaxies at magnitudes just fainter than 
our study.
ACS coverage was limited primarily to the central region of Coma, however, so we are unable to establish whether VLSBs originated at large cluster-centric distances or via tidal processes near the core.
Further ACS observations at the infall region of Coma would be useful for this purpose.
We note that compact galaxies (existing either prior to the tidal interaction or afterwards owing to a strong nuclear starburst) are more likely to survive this process and reach the cluster virial radius.

The deficit of dwarf galaxies in Coma may also result from the pre-processing of galaxies in groups located in intercluster filaments (e.g., via tidal interactions).
Recent studies of compact group galaxies (CGGs) have identified a deficit of SF galaxies with UV and IR properties that are otherwise common to a wide range of environments
\citep[albeit with more emphasis on normal galaxies;][]{Johnson2007,Tzanavaris2010,Walker2010}.
Interestingly, \cite{Walker2010} found that the properties of Coma member galaxies at the infall region studied here were the best match to the CGG population.
We note, however, that pre-processing in groups is likely a secondary effect as slightly less than 50\% of all dwarf galaxies accreted onto local galaxy clusters originate from groups \citep{McGee2009}.
Although studies have started to examine the UV properties of dwarf galaxies in groups and also filaments \citep[e.g.,][]{Haines2011,Mahajan2011}, deeper surveys are needed
to constrain the SF population at the turnover identified here.

The most likely scenario in our estimation is that the deficit of dwarf galaxies in Coma results from tidal interactions with infalling substructure outside the cluster virial radius.
Dwarf SF galaxies that survive this process are quenched via ICM processes inside the cluster.
Deep {\it GALEX} observations recently performed at the center of Coma (P.I.~R.~Smith) will place further constraints on the evolution of dwarf galaxies in Coma.
The core region also has deep HST-ACS coverage \citep{Carter2008} allowing for detailed structural information on cluster members.

\subsection{Summary}
We report on a deep {\it GALEX} imaging survey at the infall region of the Coma cluster, which for the first time has placed constraints on the Coma FUV and NUV LFs in the 
magnitude range $-14.0$$\simlt$M$_{\mathrm{UV}}$$\simlt$$-10.5$. This study reports the following results:

\begin{trivlist}

\item [$\bullet$] The Coma FUV and NUV LFs for SF galaxies show a turnover at faint magnitudes (M$_{\mathrm{UV}}$$\simgt$$-14$) that is associated
with a deficit of dwarf SF galaxies with stellar masses below M$_{*}\approx10^{8}$ $\msun$.
The same turnover is identified in UV LFs for the Virgo cluster \citep{Boselli2011}, whereas the field UV LFs \citep[e.g.,][]{Treyer2005} increase to fainter magnitudes.
This is direct evidence that dwarf SF galaxies are more efficiently processed in dense environments as compared to higher-mass galaxies.

\item [$\bullet$] The shapes of the Coma NUV LFs for dwarf passive and dwarf SF galaxies are well matched after correcting for color differences,
with both LFs showing the same turnover (dwarf passive galaxies are more numerous by a factor of $\sim$4-5).
This suggests that the recent quenching of SF galaxies via ICM processes is sufficient to build the dwarf passive population at the infall
region of Coma. The similar LFs also imply that the infall population has a deficit of dwarf SF galaxies prior to reaching the cluster virial radius, possibly due to severe tidal
disruption of dwarf galaxies along intercluster filaments.

\item [$\bullet$] There is a lack of confirmed Coma member galaxies, or gap, in a small region of the UV-optical CMDs associated with dwarf SF galaxies.
The gap is found over a narrow range of optical magnitudes at $-16.5\simlt$ M$_{r}\simlt-15.0$ corresponding to stellar masses of M$_{*}$$\simeq$10$^{7.5-8.0}$ $\msun$.
A comparison to dwarf SF galaxies observed in the field \citep{Hunter2010} indicates that the gap and the overall deficit of dwarf SF 
galaxies in Coma may reflect a lack of low surface brightness (dIrr) galaxies in Coma.

\item [$\bullet$] The Coma UV LFs were fit using both a single Schechter function and a two-component Gaussian-plus-Schechter model.
The two-component model provides a better parametrization of the UV LFs, especially for SF galaxies, and results in shallower faint-end slopes.
Fitting the total population gives faint-end slopes of $\alpha$ $\simeq$ $-1.15$ $\pm$ $0.12$ in both {\it GALEX} bands as compared to 
$\alpha$ $=-1.39$ $\pm$ 0.06 for the single Schechter model.
The two-component fits to the separate passive and star forming UV LFs were performed by aligning the Schechter component to the dwarf galaxy 
population \citep[the expected distribution for dIrr and dE galaxies; e.g.,][]{DeLapparent2003} resulting in faint-end slopes shallower than $\alpha$ $=-1$.

\item [$\bullet$] We have demonstrated that LF studies that estimate background contamination using a control field may significantly overestimate the number density and faint-end slope owing to large-scale structure behind Coma.  The spectroscopic completeness method may also overestimate the number density and 
faint-end slope (for any LF measured in a different band than used to select redshift targets) by: (1) not measuring the LF separately for blue and red galaxies, 
or (2) including galaxies fainter than the spectroscopic redshift limit.

\end{trivlist}

\acknowledgements
We thank the anonymous referee for their detailed comments and suggestions that have improved this paper.
We thank R.~Marzke and members of the Coma Hectospec team for their work on the redshift catalog, and M.~Colless for providing an updated version of the GMP catalog redshifts.
We are grateful for the LF data shared by A.~Boselli, M.~Treyer, and M.~Blanton, and assistance with the GALEV software by R.~Kotulla.
We thank T.~Heckman for commenting on an early draft and A.~Basu-Zych, P.~Tzanavaris, and B.~Lehmer for helpful science discussion.
This research was partially supported by the {\it GALEX} Cycle 2 grant 05-GALEX05-0046 (P.I.~Hornschemeier). {\it GALEX} is a NASA Small Explorer, developed 
in cooperation with the Centre National d'Etudes Spatiales of France and the Korean Ministry of Science and Technology. Funding for the creation and 
distribution of the SDSS Archive has been provided by the Alfred P. Sloan Foundation, the Participating Institutions, NASA, the NSF, DoE, Monbukagakusho, 
Max Planck Society, and the Higher Education Funding Council for England. The SDSS Web Site is http://www.sdss.org/. This study made use of the NASA
Extragalactic Database (NED) which is operated by the Jet Propulsion Laboratory (Cal Tech), under contract with NASA.

\appendix
\section{{\it GALEX} and SDSS Photometric Completeness}
{\it GALEX} completeness was addressed in Paper I as simulations were performed to measure the probability that a UV source was detected
for a given optical counterpart. We were able to recover the UV flux for 95\% of the simulated sources brighter
than NUV=24.5 and FUV=24.5 mag.

The SDSS DR6 catalog is 95\% complete for point sources to $r$=22.2 mag \citep{Adelman2008},
but this limit may be $\sim$1 mag brighter for extended galaxies owing to their lower surface brightness\footnote{www.sdss.org/dr6/products/general/completeness.html}.
We have measured the SDSS $r$-band completeness, or detection efficiency, both for Coma members and background galaxies by
matching DR6 to much deeper HST-ACS source catalogs at the center of the Coma cluster \citep[the ACS survey was performed in the F475W and F814W passbands;][]{Carter2008,Hammer2010b}.
Coma members in the ACS catalogs were identified from the spectroscopic redshift coverage (similar to the redshift coverage described in Section 2.2) and also using morphology selection \citep[see][]{Chiboucas2010};
the majority of dwarf galaxies in Coma are low surface brightness dE and dSF galaxies, with a small fraction of compact ellipticals.
In cases where SDSS did not record a detection, the $r$-band magnitude was derived from the typical SDSS/ACS colors of similar galaxies.
In Table 3, we give the SDSS $r$-band detection efficiencies for background and Coma member galaxies fainter than $r$=18 (we assume 100\% completeness at brighter magnitudes).

We confirm that the SDSS 95\% completeness limit for background galaxies is $\sim$1 mag brighter than for point sources,
and drops rapidly to only a few percent for galaxies fainter than $r$=21.5 mag.
The detection efficiency for Coma member galaxies is similar to the background galaxies but with a brighter and more gradual
decline that starts at $r$=20.5 mag. The brighter decline results from Coma dwarf galaxies that,
for a given apparent magnitude, have lower surface brightness than the average SDSS galaxy located at higher redshifts ($z$$\sim$0.1-0.2).

We have also estimated the SDSS detection efficiency for galaxies as a function of {\it GALEX} FUV and NUV
magnitudes.  This was measured in intervals of UV magnitude by summing the $r$-band detection efficiencies
for each galaxy and dividing by the total number of galaxies: DE$_{UV}$=$\sum_{i=1}^{n_{gal}}$DE$_{r}$(i)/n$_{gal}$.
We have further separated the Coma member and background samples into blue and red galaxies based on their 
UV-optical colors. These values are provided in Table 4.

\clearpage
\bibliographystyle{apj}
\bibliography{ms}
\clearpage

\clearpage

\clearpage

\clearpage

\clearpage

\clearpage


\clearpage

\clearpage

\clearpage

\clearpage

\begin{figure}[t!]
\vspace{40pt}
\centerline{\scalebox{0.92}{\rotatebox{0.0}{\includegraphics*[15,19][276,374]{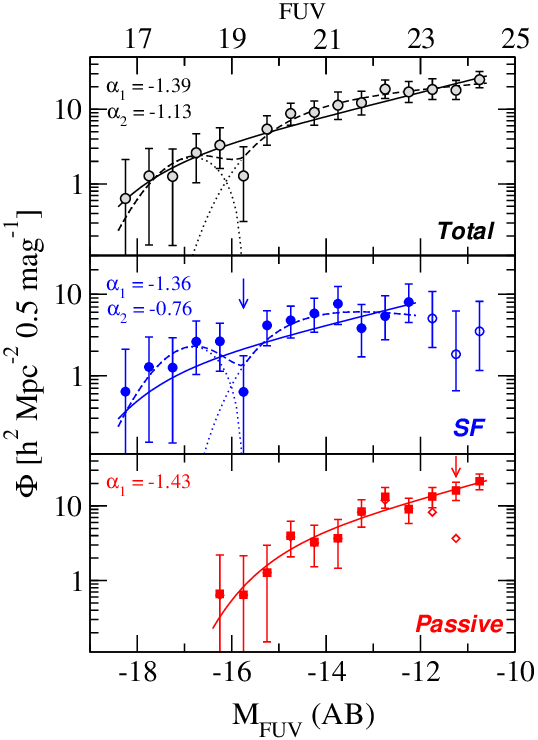}}
\includegraphics*[15,19][276,374]{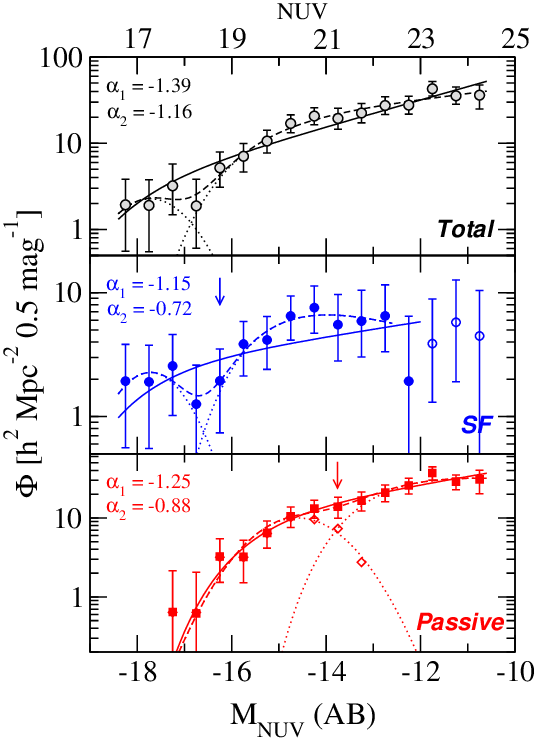}}}
\caption{\label{uvlf_fit} Fits to the Coma FUV and NUV LFs for the total galaxy population (top panels), star forming galaxies (middle panels),
and passive galaxies (bottom panels).  In each panel, the solid line shows the best-fit Schechter function.
Dotted lines show the separate components of the Gauss-plus-Schechter models with
dashed lines showing the sum of both functions. The faint-end slopes for the Schechter ($\alpha_{1}$) and Gauss-plus-Schechter ($\alpha_{2}$) fits
are listed in the upper left corner of each panel.  The full list of fitted parameters is given in Table 2.
Arrows in the bottom two panels indicate the magnitudes where dwarf galaxies (M$_{r}>-18.5$)
become the majority population; the Schechter component of the Gauss-plus-Schechter model is essentially a fit to the dwarf galaxy population for the bottom two panels.
Diamonds in the bottom panels show the LF for normal passive galaxies (M$_{r}<-18.5$). We did not fit a Gauss-plus-Schechter model to the FUV passive LF as dwarf
galaxies are the majority in only the last few magnitude intervals.
The Gauss-plus-Schechter models, in addition to providing a more physical interpretation of the SF and passive LFs (see Section 4.1.2), are a better match to the shapes of the total and SF LFs.}
\end{figure}
\clearpage


\begin{figure}[t!]
\vspace{50pt}
\centerline{\scalebox{0.54}{\rotatebox{0.0}{\includegraphics*[9,16][481,377]{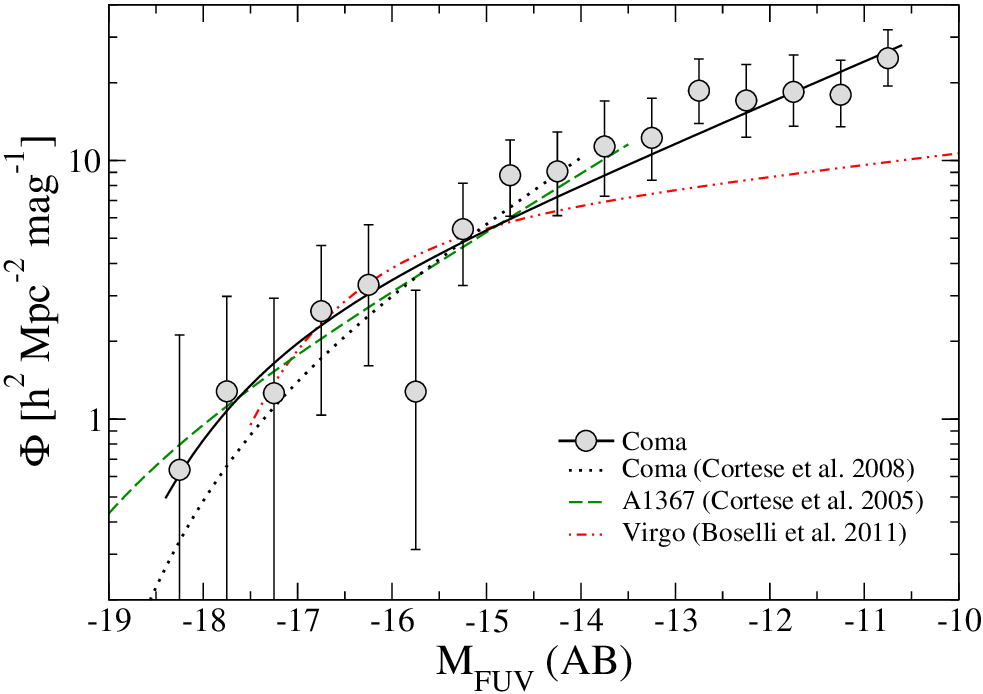}}
\includegraphics*[9,16][481,377]{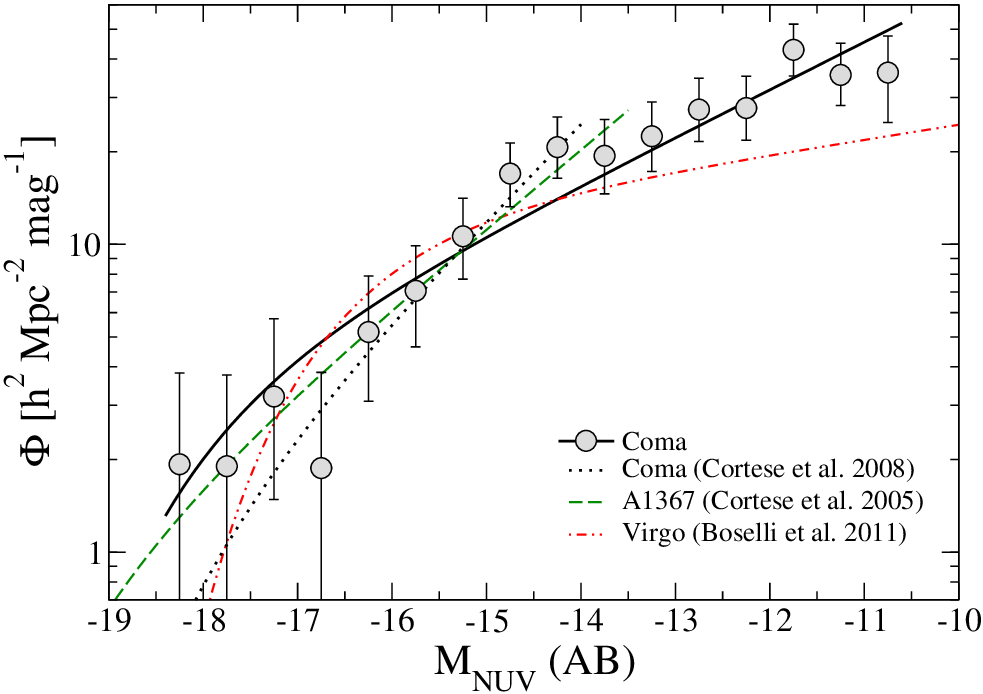}}}
\caption{\label{uvlf_other}
FUV and NUV LFs of the Coma cluster (gray filled circles; black solid line)
compared to previous {\it GALEX} LFs in Coma (black dotted), A1367 (green dashed), and Virgo (red dashed-dot).
The comparison LFs were derived from the published Schechter parameters and normalized to the integral of our Coma LF at magnitudes brighter than M$_{\mathrm{UV}}=-14$.
The shapes of the LFs are similar at bright magnitudes despite their different environments and cluster-centric coverage (see Section 4.2.1).
Although our Coma LFs agree with \cite{Cortese2008b} across magnitudes where there is overlap (M$_{\mathrm{UV}}<-14$),
our faint-end slopes are much shallower owing to a flattening of the Coma UV LFs at faint magnitudes (especially in the NUV band).
}
\end{figure}
\clearpage

\begin{figure}[t!]
\vspace{40pt}
\plottwo{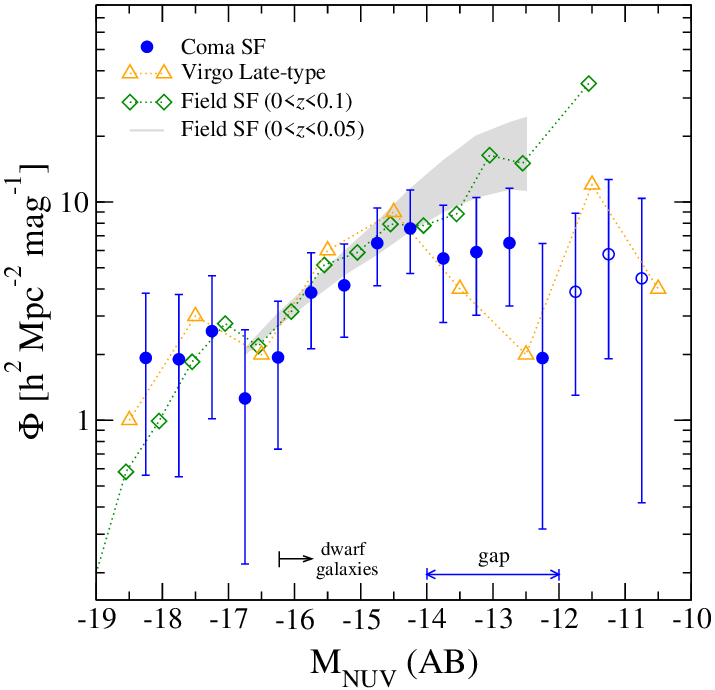}{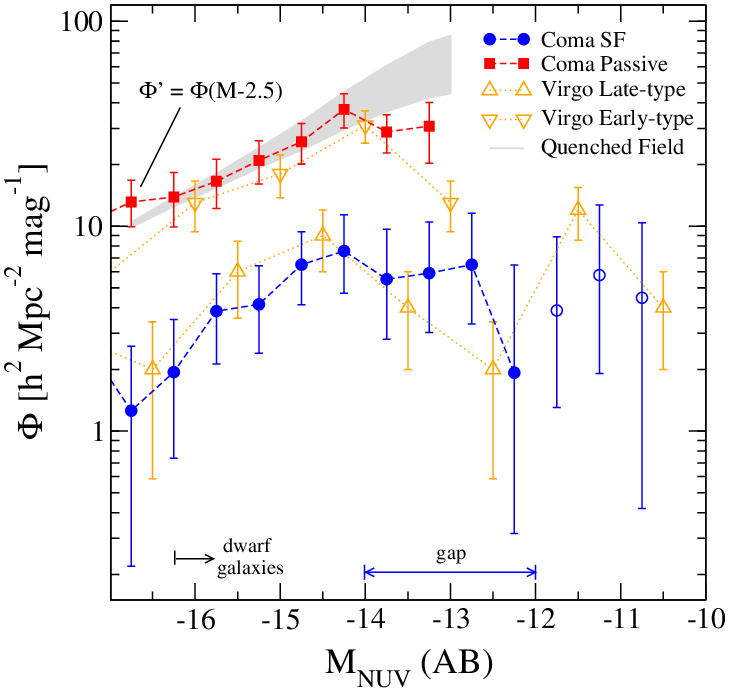}
\caption{\label{nuvlfpts}
({\it Left}) NUV LF for star forming galaxies in Coma (blue circles) as compared to the Virgo cluster \citep[orange triangles;][]{Boselli2011}
and the field environment \citep[green diamonds;][]{Treyer2005}. The gray shaded region is another field LF
that we derived from previous SDSS and {\it GALEX} LFs \citep[see Section 4.2.1;][]{Blanton2005,Wyder2007}.
The bottom of the diagram shows the magnitude ranges associated with dwarf galaxies in Coma and the gap.
The field LFs were normalized to the average of Virgo and Coma at magnitudes brighter than M$_{\mathrm{NUV}}$$=-14$.
The Virgo LF is shown at its true value.
The Coma and Virgo LFs have similar shapes and show the same turnover at M$_{\mathrm{NUV}}=-14$, whereas the field LFs continue to rise at faint magnitudes.
This suggests that external processes related to the cluster environment are more efficient in disrupting fainter (lower mass) galaxies.
({\it Right}) Same as the left panel but now including passive galaxies in Coma (red squares) and
Virgo (orange triangles - pointing down).  We have shifted the passive LFs to brighter magnitudes (2.5 mag)
to match the typical color separation of star forming and passive galaxies at the turnover.
The matched LFs follow similar shapes and show the same turnover,
suggesting that the dwarf passive population in each cluster may be built from recently quenched star forming galaxies.
The shaded region models the expected distribution of dwarf passive galaxies for a field galaxy population that is quenched but not destroyed
(see Section 4.2.1); the divergence of the passive LFs relative to this model is consistent with an infall population 
that has a turnover prior to reaching the cluster virial radius.
}
\end{figure}
\clearpage

\begin{figure}[t!]
\vspace{50pt}
\epsscale{0.6}
\plotone{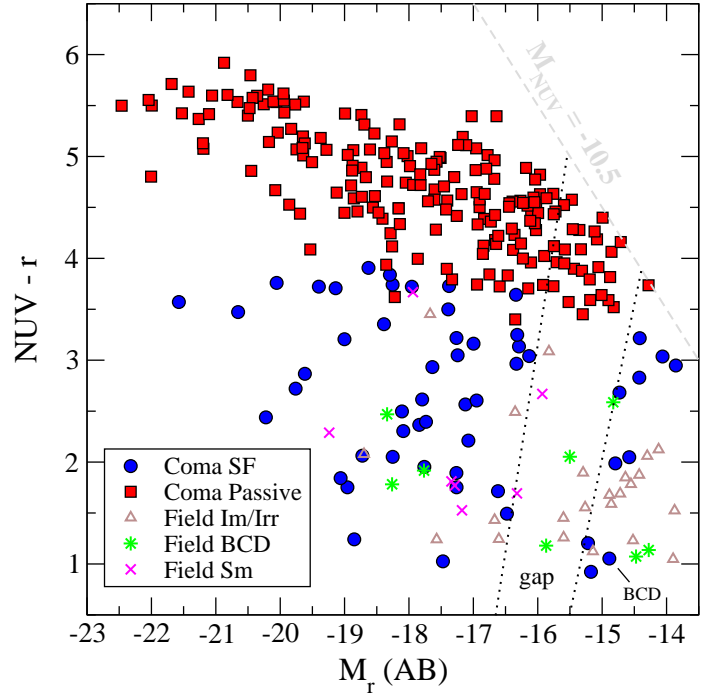}
\caption{\label{GAP} NUV-r vs.~M$_{r}$ CMD comparing Coma member galaxies and dwarf star forming galaxies located in the local 
field/group environment \citep{Hunter2010}.  Coma members are separated into passive (red squares) and star forming galaxies (blue circles); a
candidate BCD galaxy in Coma is identified in the bottom right corner of the diagram.
Field galaxies are separated by morphology as Sm (magenta Xs), BCD (green asterisks),
and Im/Irr galaxies (brown open triangles). We re-reddened the field galaxies by the internal dust prescription used in \cite{Hunter2010} and
converted their optical $V$-band to the SDSS $r$-band based on typical SEDs for these galaxies.
The gray dashed line shows the NUV magnitude limit of this study and black dotted lines enclose the gap.
Im/Irr galaxies are the majority galaxy type in the field starting at the gap (and extending to fainter magnitudes), and hence
are the likely candidate for the missing dwarf star forming galaxies in Coma.
}
\end{figure}
\clearpage

\include{tab1}
\include{tab2}
\include{tab3}
\include{tab4}

\end{document}

%% file: definitions.tex
\newcommand{\msun}{\hbox{${\mathrm{M}}_{\odot}$}}

\newcommand{\simgt}{\lower 2pt \hbox{$\, \buildrel {\scriptstyle >}\over {\scriptstyle\sim}\,$}}
\newcommand{\simlt}{\lower 2pt \hbox{$\, \buildrel {\scriptstyle <}\over {\scriptstyle\sim}\,$}}



%% file: tab1.tex
\begin{deluxetable}{lcccccc}
\tablecolumns{7}
\tabletypesize{\scriptsize}
\tablecaption{{\it GALEX} NUV and FUV LFs in Coma\label{lf_bins}}
\tablewidth{0pt}
\tablenum{1}
\tablehead{
\colhead{} & \multicolumn{3}{c}{\underline{NUV}} & \multicolumn{3}{c}{\underline{FUV}} \\
\colhead{M$_{UV}$} & \colhead{$\phi$(m)} & \colhead{$\delta_{u}\phi$(m)} &  \colhead{$\delta_{l}\phi$(m)} & \colhead{$\phi$(m)} & \colhead{$\delta_{u}\phi$(m)} & \colhead{$\delta_{l}\phi$(m)}}
\startdata
\cutinhead{Total}
-18.25 	&	1.9 	&	1.9	&	1.4	&	0.6	&	1.5	&	0.8	\\
-17.75 	&	1.9 	& 	1.9 	&	1.4	&	1.3	&	1.7	&	1.1	\\
-17.25 	&	3.2	&	2.5	&	1.7 	&	1.3	&	1.7	&	1.1	\\
-16.75 	&	1.9	&	2.0	&	1.3 	&	2.6	&	2.1	&	1.6	\\
-16.25 	&	5.2	&	2.7	&	2.1 	&	3.3	&	2.3	&	1.7	\\	
-15.75 	&	7.1 	&	2.8	&	2.4 	&	1.3	&	1.9	&	1.0	\\
-15.25 	&	10.6 	&	3.5	&	2.9 	&	5.4	&	2.7	&	2.1	\\
-14.75 	&	17.0 	&	4.4	&	3.7 	&	8.8	&	3.2	&	2.7	\\
-14.25 	&	20.7	&	5.2	&	4.3 	&	9.1	&	3.8	&	3.0	\\
-13.75	&	19.4	&	6.0	&	4.8	&	11.3	&	5.7	&	4.1	\\
-13.25  &       22.5   	&       6.5    &        5.2  	&       12.2    &	5.2	&	3.9	\\
-12.75  &       27.4 	&       7.2 	&       5.8	&       18.6    &	6.1	&	4.7	\\
-12.25  &       27.7 	&       7.5 	&       5.9	&       17.1    &	6.4	&	4.8	\\
-11.75  &       42.9   	&       9.1    &        7.6	&       18.4    &	7.2	&	4.9	\\
-11.25  &       35.5 	&       9.5 	&       7.2     &       18.0    &	6.4	&	4.5	\\
-10.75  &       36.2 	&       11.3	&       11.2    &       24.9    &	7.1	&	5.5	\\
\cutinhead{Star Forming Galaxies}
-18.25  &       1.9     &       1.9     &       1.4     &       0.6     &       1.5     &       0.8     \\
-17.75  &       1.9     &       1.9     &       1.4     &       1.3     &       1.7     &       1.1     \\
-17.25  &       2.6     &       2.0     &       1.5     &       1.3     &       1.7     &       1.1     \\
-16.75  &       1.3     &       1.3     &       1.0     &       2.6     &       2.1     &       1.6     \\
-16.25  &       1.9     &       1.6     &       1.2     &       2.6     &       1.8     &       1.5     \\
-15.75  &       3.8    	&       2.0     &       1.7     &       0.6     &       1.1     &       0.6     \\
-15.25  &       4.2    	&       2.3     &       1.8     &       4.2     &       2.1     &       1.8     \\
-14.75  &       6.5     &       2.9     &        2.3     &       4.8    	&       2.4     &       1.9     \\
-14.25  &       7.6     &       3.8     &        2.9     &       5.8    	&       3.1     &       2.4     \\
-13.75  &       5.5    	&       4.2    	&       2.7     &       7.7    &        4.8     &       3.4     \\
-13.25  &       5.9    	&       4.6    	&       2.9     &       3.8    	&       3.6     &       2.1     \\
-12.75  &       6.5     &       5.1    	&       3.2     &       5.4    	&       4.2    	&       2.6     \\
-12.25  &       1.9    	&       4.5    	&       1.6     &       8.0    &        5.3    	&       3.5     \\
-11.75  &       5.7    	&       5.6    	&       2.8    	&       5.0    	&       5.7    	&       2.8     \\
-11.25  &       6.7    	&       7.2    &        3.9    	&       1.8    	&       4.4     &       1.2     \\
-10.75  &       5.4    	&       6.3    	&       4.1    	&       3.5    	&       4.7    	&       2.3     \\
\cutinhead{Passive Galaxies}
-17.25  &       0.6     &       1.5     &       0.8     &       \nodata     &     \nodata     &       \nodata     \\
-16.75  &       0.6     &       1.4     &       0.7     &       \nodata     &     \nodata     &       \nodata     \\
-16.25  &       3.3     &       2.2     &       1.7     &       0.7     &       1.5     &       0.8     \\
-15.75  &       3.2     &       2.0     &       1.7     &       0.6     &       1.5     &       0.8     \\
-15.25  &       6.5    &       2.7     &       2.3      &       1.3     &       1.7     &       1.1     \\
-14.75  &       10.5    &       3.2     &       2.9     &       4.0     &       2.2     &       1.9     \\
-14.25  &       13.1    &       3.6     &       3.2     &       3.3     &       2.2     &       1.7     \\
-13.75  &       13.9    &       4.4     &       4.0     &       3.7     &       2.9     &       2.2     \\
-13.25  &       16.6    &       4.7     &       4.4     &       8.4     &       3.7     &       3.2     \\
-12.75  &       20.9    &       5.2     &       4.9     &       13.2    &       4.4     &       3.9     \\
-12.25  &       25.8    &       5.9     &       5.7     &       9.0     &       3.6     &       3.2     \\
-11.75  &       37.1    &       7.2    &       7.0      &       13.4    &       4.3     &       4.0     \\
-11.25  &       28.8    &       6.1    &       6.1      &       16.1    &       4.7     &       4.3     \\
-10.75  &       30.8    &       9.4    &       10.5     &       21.4    &       5.4     &       4.9     \\
\enddata
\tablecomments{{LFs} are in units of galaxies 0.5 mag$^{-1}$ Mpc$^{-2}$.}
\end{deluxetable}

%% file: tab2.tex
\begin{deluxetable}{cccccc}
\tablecolumns{6}
\tabletypesize{\scriptsize}
\tablecaption{Fitted Schechter and Gauss$+$Schechter Parameters for Coma UV LFs}
\tablewidth{0pt}
\tablenum{2}
\tablehead{
\colhead{Band} & \colhead{Model}  &  \colhead{M$_{\mathrm{Lim}}$}   & \colhead{$\phi^{*}$} & \colhead{M$^{*}$} & \colhead{$\alpha$}}
\startdata
\cutinhead{All Galaxies}
NUV	&	S	&	-10.5	&	3.2 	&	-18.5$\pm$0.5 	&	-1.39$\pm$0.04	\\
NUV     &       GS	&	-10.5   &       18.7    &       -15.8$\pm$0.4   &       -1.16$\pm$0.09  \\
\\
FUV	&	S	&	-10.5	&	1.8 	&	-18.2$\pm$0.9 	&  	-1.39$\pm$0.06	\\
FUV	&	GS	&	-10.5	&	13.4	&	-15.2$\pm$0.6	&	-1.13$\pm$0.14	\\
\cutinhead{Star Forming Galaxies}
NUV     &       S	&	-12.0	&	2.4     &       -18.5$\pm$0.5   &       -1.15$\pm$0.12	\\
NUV	&	GS	&	-12.0	&	15.2	&	-15.1$\pm$0.6	&	-0.48$\pm$0.43	\\
\\
NUV	&	S	&	-12.5	&	1.8	&	-18.5$\pm$0.5	&	-1.28$\pm$0.14	\\
NUV     &       GS	&	-12.5	&	12.5    &       -15.3$\pm$0.6   &       -0.72$\pm$0.30	\\
\\
FUV     &       S	&	-12.0   &       1.1     &       -18.2$\pm$0.9   &       -1.36$\pm$0.15	\\
FUV     &       GS	&	-12.0	&	10.9    &       -14.8$\pm$1.0   &       -0.76$\pm$0.30	\\
\cutinhead{Passive Galaxies}
NUV     &       S	&	-10.5	&	11.3    &       -15.8$\pm$0.5   &       -1.25$\pm$0.10	\\
NUV     &       GS	&	-10.5   &       46.0    &       -13.1$\pm$0.6   &       -0.88$\pm$0.30	\\
\\
FUV     &       S	&	-10.5	&	3.2     &       -15.5$\pm$1.2   &       -1.43$\pm$0.18	\\
\enddata
\tablecomments{Models are GS=Gauss+Schechter and S=Schechter, and M$_{\mathrm{Lim}}$ is the faint magnitude limit.
The fitted Schechter parameters are $\phi^{*}$ (0.5 mag$^{-1}$ Mpc$^{-2}$), M$^{*}$ (mag), and $\alpha$.
The fitted Gaussian parameters for the GS models of the NUV(FUV) LFs are: $\phi_{\mathrm{pk}}=2.3$($1.1$) 0.5 mag$^{-1}$ Mpc$^{-2}$, M$_{\mathrm{avg}}=-17.7$($-16.1$) mag, $\sigma_{\mathrm{M}}=0.76$($0.86$) mag, and skew$=0.0$($-2.3$).}
\end{deluxetable}

%% file: tab3.tex
\begin{deluxetable}{cccc}
\tablecolumns{4}
\tabletypesize{\scriptsize}
\tablecaption{SDSS Detection Efficiency for Extended Galaxies}
\tablewidth{0pt}
\tablenum{3}
\tablehead{
\colhead{{\it r}} & \colhead{Det.~Eff.} & \colhead{$\delta$$_{u}$} & \colhead{$\delta$$_{l}$}}
\startdata
\cutinhead{Background galaxies}
18.25 	&	1.00 	&	0.00	&	0.13	\\
18.75 	&	1.00 	& 	0.00 	&	0.17	\\
19.25 	&	0.88	&	0.06	&	0.10 	\\
19.75 	&	0.97	&	0.02	&	0.06 	\\
20.25 	&	0.99	&	0.01	&	0.03 	\\	
20.75 	&	0.97 	&	0.02	&	0.03 	\\
21.25	&	0.94 	&	0.02	&	0.03 	\\
21.75 	&	0.04 	&	0.00	&	0.00 	\\
22.25 	&	0.02	&	0.00	&	0.00 	\\
\cutinhead{Coma member galaxies}
18.25   &       1.00    &       0.00    &       0.17	\\
18.75   &       0.93    &       0.06    &       0.14	\\
19.25   &       1.00    &       0.00    &       0.08    \\
19.75   &       1.00    &       0.00    &       0.12    \\
20.25   &       1.00    &       0.00    &       0.08    \\
20.75   &       0.89    &       0.07    &       0.13    \\
21.25   &       0.70    &       0.10    &       0.13    \\
21.75   &       0.18    &       0.14    &       0.09    \\
22.25   &       0.23    &       0.18    &       0.12    \\
\enddata
\end{deluxetable}

%% file: tab4.tex
\begin{deluxetable}{ccccccc}
\tablecolumns{7}
\tabletypesize{\scriptsize}
\tablecaption{SDSS Detection Efficiency for Extended Galaxies in the {\it GALEX} Bands}
\tablewidth{0pt}
\tablenum{4}
\tablehead{
\colhead{} & \multicolumn{3}{c}{\underline{NUV}} & \multicolumn{3}{c}{\underline{FUV}} \\
\colhead{AB mag} & \colhead{D.E.} & \colhead{$\delta$$_{u}$} & \colhead{$\delta$$_{l}$} & \colhead{D.E.} & \colhead{$\delta$$_{u}$} & \colhead{$\delta$$_{l}$}}
\startdata
\cutinhead{Background galaxies (red)}
21.25	&	1.00 	&	0.00	&	0.02 	&	1.00	&	0.00	&	0.00	\\
21.75 	&	1.00 	&	0.00	&	0.01 	&	1.00	&	0.00	&	0.00	\\
22.25 	&	1.00	&	0.00	&	0.02 	&	1.00	&	0.00	&	0.00	\\
22.75	&	1.00	&	0.00	&	0.02	&	1.00	&	0.00	&	0.00	\\
23.25	&	0.99	&	0.01	&	0.02	&	1.00	&	0.00	&	0.01	\\
23.75	&	0.98	&	0.01	&	0.02	&	1.00	&	0.00	&	0.04	\\
24.25	&	0.98	&	0.01	&	0.02	&	1.00	&	0.00	&	0.02	\\
\cutinhead{Background galaxies (blue)}
19.25   &       1.00    &       0.00    &       0.02    &	1.00    &       0.00    &       0.01	\\
19.75   &       1.00    &       0.00    &       0.01    &	1.00    &       0.00    &       0.01	\\
20.25   &       1.00    &       0.00    &       0.01    &	1.00    &       0.00    &       0.01	\\
20.75   &       1.00    &       0.00    &       0.01    &	0.99    &       0.00    &       0.01	\\
21.25   &       1.00    &       0.00    &       0.01    &	0.99    &       0.00    &       0.01	\\
21.75   &       1.00    &       0.00    &       0.00    &	0.98	&	0.00	&	0.01	\\
22.25   &       1.00    &       0.00    &       0.00    &	0.98    &       0.00    &       0.00	\\
22.75   &       1.00    &       0.00    &       0.00    &	0.97    &       0.00    &       0.00	\\
23.25   &       0.99    &       0.00    &       0.00    &	0.97    &       0.00    &       0.01	\\
23.75   &       0.98    &       0.00    &       0.01    &	0.97	&	0.00	&	0.01	\\
24.25   &       0.98    &       0.01    &       0.02    &	0.96	&	0.00	&	0.01	\\
\cutinhead{Coma member galaxies (red)}
21.75   &       1.00    &       0.00    &       0.00    &	1.00	&	0.00	&	0.00	\\
22.25   &       1.00    &       0.00    &       0.01    &       1.00    &       0.00    &       0.00    \\
22.75   &       0.98    &       0.01    &       0.01    &       1.00    &       0.00    &       0.00    \\
23.25   &       0.97    &       0.01    &       0.01    &       1.00    &       0.00    &       0.00    \\
23.75   &       0.99    &       0.00    &       0.02    &       1.00    &       0.00    &       0.00    \\
24.25   &       0.98    &       0.01    &       0.04    &       1.00    &       0.00    &       0.01    \\
\cutinhead{Coma member galaxies (blue)}
19.25   &       1.00    &       0.00    &       0.01    &	1.00	&	0.00	&	0.01	\\
19.75   &       1.00    &       0.00    &       0.01    &	1.00	&	0.00	&	0.01	\\
20.25   &       1.00    &       0.00    &       0.01    &	0.99	&	0.01	&	0.01	\\
20.75   &       1.00    &       0.00    &       0.03    &	1.00	&	0.00	&	0.02	\\
21.25   &       1.00    &       0.00    &       0.03    &	1.00	&	0.00	&	0.03	\\
21.75   &       0.93    &       0.03    &       0.03    &	1.00	&	0.00	&	0.02	\\
22.25   &       0.97    &       0.02    &       0.03    &	1.00	&	0.00	&	0.03	\\
22.75   &       1.00    &       0.00    &       0.08    &	0.96	&	0.02	&	0.03	\\
23.25   &       0.94    &       0.04    &       0.09    &	0.93	&	0.04	&	0.04	\\
23.75   &       0.89    &       0.05    &       0.09    &	0.95	&	0.04	&	0.06	\\
24.25   &       0.70    &       0.10    &       0.13    &	0.94	&	0.04	&	0.08	\\
\enddata
\tablecomments{Brighter magnitude intervals are assumed to have perfect coverage.
The detection efficiencies for background galaxies are valid at redshifts below $z\approx0.2$.
Galaxies were classified as red/blue separately for each band based on apparent UV-optical colors: FUV$-r=5.5$ and NUV$-r=-0.14\times\mathrm{NUV}+6.88$ are the dividing lines.
The blue galaxy sample is incomplete at magnitudes fainter than m$_{\mathrm{UV}}\sim22.0 (23.0)$ for background (Coma) galaxies, respectively, 
owing to the optical photometry limit of our catalog;
the detection efficiencies therefore apply only to regions of the UV-optical CMD that have detections (see Figure \ref{uvcmd}).}
\end{deluxetable}